\documentstyle[twocolumn,eqsecnum,aps,epsf]{revtex}

\newcommand{\ztb}{\bar{Z}_{2+}}
\newcommand{\zsb}{\bar{Z}_{6+}}
\newcommand{\phib}{\bar{\Phi}_+}
\newcommand{\Cb}{\bar{C}}

\begin{document}


\title{The stability of the shell of D6-D2 branes in
a ${\cal N}=2$ supergravity solution}

\author{Kengo Maeda
\thanks{electronic mail:maeda@cosmic.physics.ucsb.edu}}
\address{Department of Physics, University of California,
Santa Barbara, CA. 93106, USA}

\author{Takashi Torii
\thanks{electronic mail:torii@resceu.s.u-tokyo.ac.jp}}
\address{Research Center for the Early Universe,
University of Tokyo,
Bunkyo-ku, Tokyo, 113-0033, Japan, and \\
Advanced Research Institute for Science and Engineering,
Waseda University,
Shinjuku-ku, Tokyo 169-8555, Japan}

\author{Makoto Narita
\thanks{electronic mail:narita@gravity.phys.waseda.ac.jp}}
\address{Advanced Research Institute for Science and Engineering,
Waseda University,
Shinjuku-ku, Tokyo 169-8555, Japan}

\author{Shigeaki Yahikozawa
\thanks{electronic mail:yahiko@rikkyo.ac.jp}}
\address{Department of Physics, Rikkyo University, Toshima-ku,
Tokyo 171-8501, Japan}

\date{\today}

\maketitle
\begin{abstract}
{\small The stability of the shell of wrapped D6-branes on K3 is
investigated from the point of view of supergravity.
We first construct an effective energy-momentum tensor for
the shell under the reasonable conditions and show that
supersymmetric solutions satisfy Israel's junction conditions
at arbitrary radius of the shell.
Next we study the perturbation of the whole system including
the self-gravity of the shell.
It is found that in spite of the existence of wrapped D6-branes
with negative tension, there is no eigenmode whose frequencies
of the shell and the fields are imaginary numbers, at any radius of
the shell.
Furthermore, when the radius of the shell is less than the
enhan\c{c}on radius, resonances are produced, and this indicates
a kind of ``instability'' of the system.
This can even classically explain why the shell is constructed
at the enhan\c{c}on radius.}
\end{abstract}

\pacs {04.70.-s, 04.70.Bw, 04.20.Jb, 95.30.Sf}


\vskip2pc]

\setcounter{footnote}{1}
\renewcommand{\thefootnote}{\fnsymbol{footnote}}

\section{Introduction}
\label{sec.I}

One of the central issues in quantum gravity is the resolution
of singularities, i.e., the initial singularity in our universe
or a central singularity in black holes.
It is expected that string theory, which is one of the most
prominent theories including the quantum gravity, may resolve
the issue of such singularities.
{}From the general relativity theory point of view, the
appearance of a naked singularity is a serious problem because
we cannot predict what happens in the future.
Therefore, it is particularly important to study whether
the naked singularities are excised in string theory or not.
An interesting example of the naked singularity is
{\it repulson singularity}, which appears in supersymmetric
solutions with spherical symmetry in supergravity
theory~\cite{B,KL,CY}.
Kallosh and Linde~\cite{KL} showed that all test
particles, either massless or massive, cannot touch
the {\it repulson singularity}.
This particular type of singularity
is a {\it globally} naked singularity and the weak cosmic
censorship~\cite{P} seems to be violated.
This is essentially different from the extreme
Reissner-Nordstr\"{o}m black hole where a {\it locally}
naked singularity exists inside the event horizon.

Recently, Johnson, Peet and Polchinski
analyzed the motion of a D6-brane probe wrapped on
a K3 manifold in the type IIA  supergravity solution
and proposed the {\it enhan\c{c}on mechanism} which excises
the repulson singularity~\cite{JPP,J}.
The higher curvature corrections on the world volume
of D6-brane induce a negative D2-charge and a
negative D2-tension~(see Refs.~\cite{BVS,GHM,DJM,BBG} for details),
and the total tension of
the wrapped D6-brane probe vanishes at a special radius $r=r_e$ larger
than the radius of the repulson singularity $|r_2|$.
This fact led the proposal that gauge symmetry is enhanced at
the radius $r=r_e$, which is called the enhan\c{c}on radius,
and that a shell consisted of the wrapped D6-branes is located
at the enhan\c{c}on radius.
As a result, the original repulson singularity can be removed
due to the flat geometry for $r<r_e$.

Since there is no precise dual theory to this type of
supergravity theory at the present stage, this motivates
us to investigate the picture of the
enhan\c{c}on mechanism from supergravity side more deeply.
It is especially interesting to examine whether or not the
enhan\c{c}on geometry is a solution of the supergravity
theory including the self-gravity of the branes,
and whether the shell of the wrapped D6-branes is
stable or collapses by the gravitational effect.

So, in this paper, we first construct an effective energy-momentum
tensor of the shell consisted of the wrapped D6-branes
and study the  static solutions by using Israel's junction
condition~\cite{Israel}.
Next we study the spherically symmetric
perturbation of the whole system and
analyze the stability of the obtained solution.
By numerical analysis, it is found that there is no eigenmode
whose frequency $\omega$ of the shell and the fields is imaginary.
To our surprise, although the tension of the wrapped D6-brane
is negative in the region $r<r_e$, the shell of the
branes never has eigenmodes with $\omega^2<0$
if we take into account the interactions between the shell and
the fields, which include the self-gravity of the shell.

Furthermore, it is shown that when the radius of the shell is
less than $r_e$, resonances are caused at proper oscillation
frequencies of the shell, namely the system becomes ``unstable"
in this region.
On the other hand, when the radius of the shell is greater than
or equal to $r_e$, the system is really stable,
that is, there is no eigenmode with $\omega^2<0$
and no resonance is produced.
These facts indicate that it may be able to explain even
classically the dynamical reason why the shell is constructed
at the enhan\c{c}on radius.

The outline of this paper is as follows.
In Sec.~\ref{sec.II}, we briefly review the D6-D2 brane
solution and show that the velocity of the D6-brane probe
reaches the speed of light at the enhan\c{c}on radius.
In Sec.~\ref{sec.III}, we derive the effective energy-momentum tensor
of the shell of the D6-D2 branes
under the reasonable conditions, and give the energy conservation law
and the dilaton equation on the Gaussian normal coordinates.
In Sec.~\ref{sec.IV}, we show ``the enhan\c{c}on geometry'' with the
shell at any radius is the solution
of the supergravity based on the Israel's junction condition.
In Sec.~\ref{sec.V}, we first derive perturbed equations of the
gravitational and matter fields inside the shell located
at arbitrary radius. Next, we derive perturbed junction
conditions of the fields on the shell, the equation
of motion of the shell and perturbed equations of fields
outside the shell.
Finally, using numerical analysis, we study the stability
of the shell and the fields.
Sec.~\ref{sec.VI} is devoted to conclusion and discussion.
Throughout this paper, we set $\alpha'=1$.

\section{Enhan\c{c}on geometry and the relativistic
motion of a D6-brane}
\label{sec.II}

The low-energy IIA supergravity action is given by
\begin{eqnarray}
\label{S-action}
S^{\rm sugra}_{\rm st}&=&\frac{1}{2{\kappa_{10}}^{2}}
\int d^{10}x \sqrt{-\hat{G}}
\biggl\{ e^{-2\Phi}\,\Bigl[\hat{R}+4\,(\hat{\nabla}\Phi)^2
\nonumber \\
&&\;\;\;\;\;\;
-\frac{1}{2}\bigl|\hat{H}_{(3)}\bigl|^2\Bigl]
-\frac{1}{2}\bigl|\hat{\tilde{F}}_{(4)}\bigl|^2
-\frac{1}{2}\bigl|\hat{F}_{(8)}\bigl|^2
\nonumber \\
&&\;\;\;\;\;\;
-\frac{1}{2\sqrt{-\hat{G}}} B_{(2)}\wedge{F}_{(4)}\wedge{F}_{(4)}
\biggl\},
\end{eqnarray}
where $\hat{G}_{MN}$ is the metric in the string frame,
$\Phi$ is the dilaton field.
$H_{(3)}$ is the field strength of the two-rank anti-symmetric
tensor field $B_{(2)}$.
$F_{(p+2)}$ is expressed by the (p+1)-form Ramond-Ramond (R-R)
field $C_{(p+1)}$ as $F_{(p+2)}=dC_{(p+1)}$ and
$\tilde{F}_{(4)}=dC_{(3)}+C_{(1)}\wedge dB_{(2)}$,
where $C_{(7)}$ is the dual field of $C_{(1)}$.
The hats on the fields and the differential denote that
the contractions are done with the string metric $\hat{G}_{MN}$.

Let us consider a D6-D2 system in which there are
N D2-branes and N D6-branes.
The supersymmetric solution of the system is
\begin{eqnarray}
\label{eq-solution}
d\hat{s}^2&=&\hat{G}_{MN}dx^Mdx^N
\nonumber \\
&=&Z_2^{-\frac{1}{2}}Z_6^{-\frac{1}{2}}\eta_{\mu\nu}
\,dx^\mu dx^\nu
+Z_2^{\frac{1}{2}}Z_6^{\frac{1}{2}}\delta_{ij}\,dx^i dx^j
\nonumber \\
&&+V^{\frac{1}{2}}Z_2^{\frac{1}{2}}Z_6^{-\frac{1}{2}}
G^{(K3)}_{pq}dx^p dx^q,
\\
\nonumber \\
e^{2\Phi}&=&g^2 Z_2^{\frac{1}{2}}Z_6^{-\frac{3}{2}}, \\
\nonumber \\
C_{(3)}&=&(g Z_2)^{-1}dx^0\wedge dx^4\wedge dx^5, \\
\nonumber \\
C_{(7)}&=&V (g Z_6)^{-1}dx^0\wedge dx^4\wedge dx^5\wedge
dV_{(K3)},
\end{eqnarray}
where $\mu,\,\nu=(0,4,5)$, $i,\,j=(1,2,3)$, and $p,\,q=(6, 7, 8, 9)$.
The $G^{(K3)}_{pq}$ denotes the metric of the K3 manifold
with unit volume\footnote{Although we cannot write
down the metric explicitly, it is well known that the Ricci
curvature is zero.
For later convenience, here, we shall formally denote
the metric as $G^{(K3)}_{pq}dx^pdx^q$}.
The D2-branes are aligned along the $4,5$ directions, while the
D6-branes are aligned to the $4,5,6,7,8,9$ directions.

The harmonic functions are expressed as
\begin{eqnarray}
Z_2=1+\frac{r_2}{r}, \qquad Z_6=1+\frac{r_6}{r},
\end{eqnarray}
where $r:=\sqrt{x^i x^i}$ and
\begin{eqnarray}
r_2:=-\frac{gNV_* }{2V}, \qquad r_6:=\frac{gN}{2},
\end{eqnarray}
$dV_{(K3)}$ is the volume form of the K3 manifold of unit
volume.
To keep the tension of a wrapped D6-brane, which will be
investigated in detail later, positive at infinity,
we assume that $V>V_*:= (2\pi)^4$.

Because of the negative charge of the D2-branes,
there is a repulson singularity at $r=|r_2|$,
where Ricci scalar curvature diverges as $(r-|r_2|)^{-3/4}$.
To investigate the causal structure of this geometry with the
repulson singularity, let us solve a radial null
geodesic in this solution. By using Eq.~(\ref{eq-solution}),
the future-ingoing null geodesic behaves near the singularity as
\begin{eqnarray}
\label{eq-null}
t=-\int^r (Z_2\,Z_6)^{\frac{1}{2}}dr
\propto (r+r_2)^{\frac{3}{2}},
\end{eqnarray}
where $t:= x^0$.
This means that $t$ is finite when $r\to |r_2|$ and hence the
repulson singularity is a timelike singularity,
as depicted in Fig.~1. To resolve the repulson singularity,
Johnson, Peet, and Polchinski have proposed the enhan\c{c}on
geometry which is flat inside the unique enhan\c{c}on
radius ~\cite{JPP}.

\begin{figure}
 \centerline{\epsfxsize=14.0cm \epsfbox{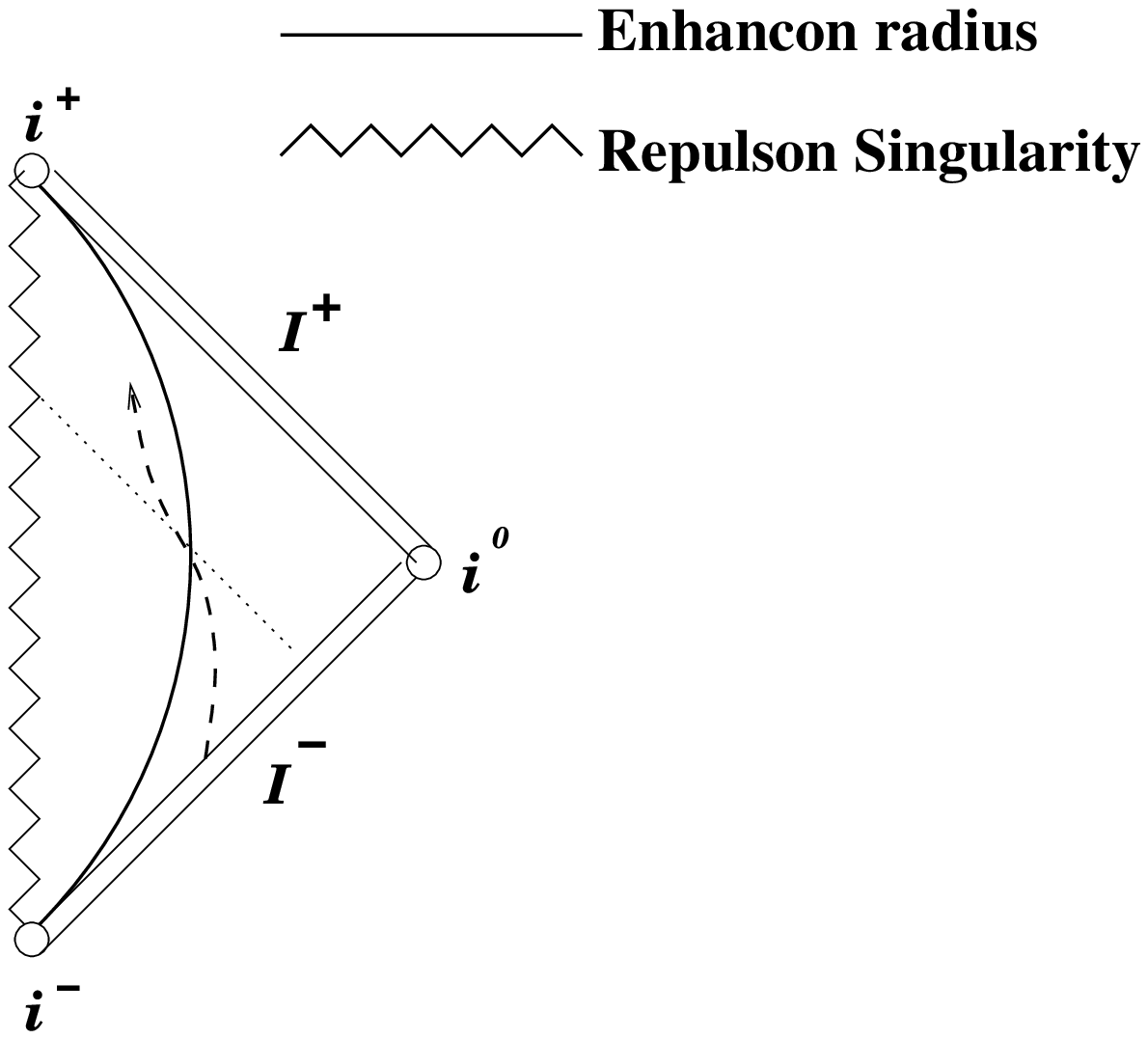}}
      \caption{The Penrose diagram of a D6-D2 supergravity
solution is shown. The dashed line represents an orbit of a
D6-brane wrapped on the K3, while the dotted line represents
the orbit of an ingoing-radial null geodesic.}
          \protect
\label{d2d6-eps}
\end{figure}

If we ignore higher order corrections including the curvature
corrections, the action of a Dp-brane is given by
\begin{eqnarray}
\label{Interaction-1}
S^{Dp}_{\rm st}&=&-\mu_p\int d^{p+1}\xi\,e^{-\Phi}
\,\Bigl[-\det(\hat{g}_{ab}+B_{ab}+2\pi F_{ab})
\Bigr]^{\frac{1}{2}}
\nonumber \\
&&+\mu_p\int_{p+1}e^{(B_{(2)}+2\pi F)}\wedge
\sum_{q}C_{(q)},
\end{eqnarray}
where $\mu_p=(2\pi)^{-p}$ is the charge of the Dp-brane and
$\hat{g}_{ab}$ is the pull-back of the string metric
$\hat{G}_{MN}$:
\begin{eqnarray}
\label{pullback}
\hat{g}_{ab}:=\frac{\partial x^M}{\partial \xi^a}
\frac{\partial x^N}{\partial \xi^b}\hat{G}_{MN}.
\end{eqnarray}
$B_{ab}$ is the component of the spacetime NS-NS fields
$B_{(2)}$ parallel to the Dp-brane and $F_{ab}$ (and $F$) is the
field strength of the gauge field living on the Dp-brane.
In this paper, we consider the D6-D2 brane system with
vanishing $B_{(2)}$ and $F_{ab}$, so that the action becomes
\begin{eqnarray}
\label{Interaction-2}
S^{Dp}_{\rm st}=&&-\mu_p\int d^{p+1}\xi\,e^{-\Phi}
\,\bigl(-\det \hat{g}_{ab}\bigr)^{\frac{1}{2}}
\nonumber \\
&&
+\mu_p\int d^{p+1}\xi\,C_{(p+1)}.
\end{eqnarray}

Now, consider the motion of a wrapped D6-brane on the K3
manifold.
If we take the higher-curvature
correction terms in the K3 to the action of the D6-brane
into account,
a D2-brane with a negative D2-charge is induced
(see~\cite{BVS,GHM,DJM,BBG} for details).
Hence, the total action of the wrapped D6-brane on the K3
becomes $S^{D6}_{\rm st}-S^{D2}_{\rm st}$.
After integration of the wrapped coordinates on the K3,
we obtain the effective action of the brane which is just
a membrane:
\begin{eqnarray}
\label{eq-D6action}
S^{D6\mbox{-}D2}_{\rm st}&=&-\mu_6 \int d^3\xi\,
e^{-\Phi(r)}(V(r)-V_*)
(-\det g_{ab})^{\frac{1}{2}} \nonumber \\
&&+\frac{\mu_{6}}{g}\int d^3\xi\,
\left(\frac{V}{Z_6}-\frac{V_*}{Z_2} \right),
\end{eqnarray}
where $V(r)$ is defined by $V(r):= VZ_2/Z_6$.
Note that the tension of the brane vanishes at the enhan\c{c}on
radius $r_e$ satisfying the following relation:
\begin{eqnarray}
\frac{V}{Z_6(r_e)}=\frac{V_*}{Z_2(r_e)}.
\end{eqnarray}
The explicit form of the enhan\c{c}on radius is
\begin{eqnarray}
\label{eq-D6action}
r_e=\frac{2V}{V - V_*}|r_2|.
\end{eqnarray}

If we take a static gauge\footnote{Since D2 brane
is a compact surface,
the space of $x^{4}$ and $x^{5}$ should be also
compact in the static gauge, to be exact. Here,
we take this space as two dimensional torus of
unit volume.}
\begin{eqnarray}
\label{eq-sgauge}
x^0=t=\xi^0, \qquad x^4=\xi^4, \qquad x^5=\xi^5,
\end{eqnarray}
the reduced Lagrangian density ${\cal L}$ is
\begin{equation}
\label{eq-rLag}
{\cal L}=\frac{\mu_6}{g}\left(\frac{V}{Z_6}-
\frac{V_*}{Z_2} \right)\left[1-
(1-Z_2Z_6\,\dot{x}^i\dot{x}^i)
^{\frac{1}{2}}\right],
\end{equation}
where dots denote the differentials with respect to $t$.

The energy conservation law gives us the following
equation
\begin{eqnarray}
\label{eq-rmotion}
\left(\frac{V}{Z_6}-\frac{V_*}{Z_2} \right)
\left[\left(1-Z_2Z_6\,\frac{dx^i}{dt}\frac{dx^i}{dt}
\right)^{-\frac{1}{2}}-1\right]
=E,
\end{eqnarray}
where $E$ is a constant.
If the probe has an ingoing-velocity at
$r=r_i\gg r_e$,
$E$ is strictly positive. If the brane probe stops
at a radius $r_m>r_e$, then the left hand side would be
zero at $r=r_m$. This leads to contradiction because $E$
is positive.
Hence the brane probe cannot stop outside of the
enhan\c{c}on radius $r_e$.
At the enhan\c{c}on radius, the first term in the square bracket
should diverge infinitely because $V/Z_6-V_*/Z_2=0$
at $r=r_e$. This means that the velocity of the brane
probe reaches the speed of light at the enhan\c{c}on
radius~(See Fig.~\ref{d2d6-eps})!

It seems strange at first sight that according to the picture
by Johnson, Peet and Polchinski in Ref.~\cite{JPP},
the branes should collect near the enhan\c{c}on radius and
that they cannot leave the location of the enhan\c{c}on radius,
while we have just seen, the probe discussed above moves at a
speed of light near the enhan\c{c}on radius.
Hence we need further analysis about this enhan\c{c}on
radius and the mechanism beyond the probe approximation.
In next sections,
we study the whole system including the self-gravity of the
branes and the self-interactions between the fields and the
branes.

\section{The energy-momentum tensor of the D6-D2
branes shell}
\label{sec.III}

By the conformal transformation such as
$G_{MN}=e^{-\Phi/2}\hat{G}_{MN}$, we obtain the
low-energy effective action in the Einstein frame:
\begin{eqnarray}
\label{E-action}
S=&&\frac{1}{2{\kappa_{10}}^2}\int d^{10}x\sqrt{-G}
\biggl[R-\frac{1}{2}(\nabla\Phi)^2
-\frac{e^{\frac{1}{2}\Phi}}{2}\,|F_{(4)}|^2
\nonumber \\
&&\;\;\;\;\;\;
-\frac{e^{-\frac{3}{2}\Phi}}{2}\,|F_{(8)}|^2\biggr]
+\sum_{i=1}^N (S^{D6}_{(i)}-S^{D2}_{(i)}),
\nonumber \\
\end{eqnarray}
where $G_{MN}$ is the Einstein metric and
all the contractions are done with $G_{MN}$.
$i$ represents the numbering of
each wrapped D6-brane of which a shell
consists.
In~\cite{Talk,JMPR}, the energy-momentum tensor
is obtained when the shell is fixed at a radius. In this section we
will obtain the tensor by the following covariant way so that it can
be applied to more general cases.

In general relativity, it is well known that there
is the following
local~(Gaussian normal) coordinates
\begin{eqnarray}
\label{E-gaussian}
ds^2=d\chi^2+q_{ab}\,d\zeta^a d\zeta^b, \qquad (a,b=0,2,..,9)
\end{eqnarray}
in the neighborhood of the D6-D2 branes shell
such that it moves along $\chi=\chi_0$ constant.

Here after, we shall impose spherical symmetry for
simplicity. Hence, we can choose the Gaussian normal
coordinates $(\chi,\zeta^a)$ as $\chi=\chi(r,t)$ and
$\zeta^0=\zeta^0(r,t)$, while the other coordinates are
same as $\theta$, $\phi$, $x^m~(m=4,5)$, $x^p$.
The nine-dimensional embedded metric is rewritten by
\begin{eqnarray}
\label{9-metric}
{ds}^2|_{\chi=\mbox{const.}}& =&
q_{ab}\,d\zeta^a d\zeta^b
\nonumber \\
&=&-q_{00}(d\zeta^0)^2+
G_{\theta\theta}d\theta^2+G_{\phi\phi}d\phi^2
\nonumber \\
&&+G_{mn}dx^m dx^n+G_{pq}dx^p dx^q.
\end{eqnarray}

To derive the effective energy-momentum tensor coming from
the wrapped D6-branes, we shall impose the following conditions:
\begin{enumerate}
\item we can take a static gauge such as
$\xi^0=\zeta^0$,
$\xi^m=x^m$ and $\xi^p=x^p$ for each D6 brane and
$\xi^0=\zeta^0$ and $\xi^m=x^m$ for each D2 brane.
\item each D6 (D2) brane moves along the radial direction only.
\item each D6 (D2) brane is distributed uniformly on the shell.
\end{enumerate}
The condition 2 implies, in other words, that
each D6-D2 brane moves
on the $\chi=\chi_0$,~$\zeta^a=\mbox{const}~(a\neq 0)$.

In order to see the way how to derive the effective
energy-momentum tensor of the D6-D2 brane shell, it
would be enough to discuss $\sum_{i=1}^N S^{D6}_{(i)}$ only.
Using the conditions 1 and 2,
the action can be reduced as follows:
\begin{eqnarray}
\label{S6}
S_6&:=&\sum_{i=1}^N (S^{D6}_{(i)}-\mu_6\int d^7\xi {C_{(7)}|}_i)
\nonumber \\
&=&-\mu_6\sum_{i=1}^N\int d^7\zeta\,e^{\frac{3}{4}\Phi}
\left(-\frac{\partial x^M}{\partial \xi^0}
\frac{\partial x^N}{\partial \xi^0}G_{MN} \right)^{\frac{1}{2}}
\biggl|_i
\nonumber \\
&& \;\;\;\;\;\;\;\;\;\;\;\;\;\;\;
\times \bigl(\det G_{mn}\cdot\det G_{pq}\bigr)^{\frac{1}{2}}
\nonumber \\ \nonumber \\
&=& -N\mu_6\int d^7\zeta\,e^{\frac{3}{4}\Phi}
\bigl(-q_{00}\cdot
\det G_{mn}\cdot\det G_{pq}\bigr)^{\frac{1}{2}}.
\nonumber \\
\end{eqnarray}
Because of the condition~3, each D6-brane is distributed
uniformly over the two-sphere ($\theta,\phi$). Introducing
the number density $\rho_6$ satisfying
\begin{eqnarray}
\label{conservation1}
\int d\theta d\phi\,(G_{\theta\theta}
G_{\phi\phi})^{\frac{1}{2}}
\,\rho_6=N,
\end{eqnarray}
we can rewrite Eq.~(\ref{S6}) as the following
ten-dimensional form:
\begin{eqnarray}
S_6&=&-\mu_6\int d^9\zeta\,e^{\frac{3}{4}\Phi}
(-\det q_{ab})^{\frac{1}{2}}\rho_6 \nonumber \\
&=&-\mu_6\int dx^{10}\,\delta(\chi-\chi_0)\,
e^{\frac{3}{4}\Phi}(-\det G)^{\frac{1}{2}}\rho_6.
\end{eqnarray}
Note that $\rho_6$ is a function of $\zeta^0$ only.
As concerns the subspace $(x^m,x^p)$, we can easily
calculate the energy-momentum tensor by using the usual
formula
\begin{eqnarray}
\label{formula}
T_{ab}=-\frac{2}{\sqrt{-q}}
\frac{\delta(\sqrt{-q}{\cal L}_m)}
{\delta q^{ab}},
\end{eqnarray}
where ${\cal L}_m$ is a matter Lagrangian.
In the subspace $(\zeta^0,\theta,\phi)$, however,
we should impose the constraint~(\ref{conservation1})
when we derive $T_{ab}$ by using Eq.~(\ref{formula}).
Let us define the three-dimensional unit timelike flow
vector along each D6-brane
by $V^{\alpha}=(V^0,V^\theta,V^\phi)
:=(\sqrt{-q^{00}},0,0)$. Then, in
terms of the current vector $J^a=\rho_6\,V^a$,
Eq.~(\ref{conservation1}) is rewritten in the following
covariant form:
\begin{eqnarray}
\label{conservation2}
{J^{\alpha}}_{;{\alpha}}=
\frac{1}{\sqrt{-q_{\beta\gamma}}}
\frac{\partial}{\partial x^{\alpha}}
(\sqrt{-q_{\beta\gamma}}J^{\alpha})=0,
\end{eqnarray}
where $;$ denotes covariant derivative with respect to
the subspace $x^{\alpha}=(\zeta^0, \theta,\phi)$. So,
as seen on page 70 of Ref.~\cite{HE},
\begin{eqnarray}
\label{conservation3}
\delta\rho_6=\frac{\rho_6}{2}
(V_{\alpha}V_{\beta}+q_{\alpha\beta})\,
\delta q^{\alpha\beta}.
\end{eqnarray}
Therefore, combining with
$\delta\sqrt{-q}/\delta q^{\alpha\beta}
=-\sqrt{-q}\,\delta q_{\alpha\beta}/2$,
we get the tensor components as
\begin{eqnarray}
\label{Stensor1}
T^{D6}_{ab}=
\left\{
\begin{array}{ll}
\displaystyle{\mu_6\,e^{\frac{3\Phi}{4}}}
\rho_6\,V_aV_b\;\delta(\chi-\chi_0),
&
(a,b=0,\theta,\phi)
\vspace{12pt}
\\
-\displaystyle{\mu_6\, e^{\frac{3\Phi}{4}}}
\rho_6\,G_{ab}\;\delta(\chi-\chi_0),
&
(a,b=m,p)
\vspace{12pt}
\\
0.
&
({\rm otherwise})
\end{array}
\right.
\nonumber
\end{eqnarray}
\vspace{-18pt}
\begin{eqnarray}
\end{eqnarray}

Just like the case of D6 branes, $\sum_{i=1}^N S^{D2}_{(i)}$
can be also
reduced as
\begin{eqnarray}
\label{S2}
S^{D2}&:=&-\sum_{i=1}^N (S^{D2}_{(i)}-\mu_2\int d^3\xi {C_{(3)}|}_i)
\nonumber \\
& =&\mu_2\int d^{10}x\,\delta(\chi-\chi_0)\,
e^{-\frac{\Phi}{4}}\,(-\det G)^{\frac{1}{2}}\,\rho_2,
\end{eqnarray}
where $\rho_2$ is a number density over the six-dimensional space,
$K3\times S^2$. It satisfies a conservation
law as
\begin{eqnarray}
\label{conservation4}
\int d^p xd\theta d\phi
\,(G_{\theta\theta}G_{\phi\phi}\,\det G_{pq})^{\frac{1}{2}}
\,\rho_2=N.
\end{eqnarray}
Defining the seven-dimensional timelike
vector $U^{\beta}=(U^0,U^\theta,U^\phi,U^p)
:=(\sqrt{-q^{00}},0,\cdots,0)$,
the effective energy-momentum tensor is
\begin{eqnarray}
\label{Stensor3}
T^{D2}_{ab}=
\left\{
\begin{array}{ll}
-
\displaystyle{\frac{\mu_2}{V^{\ast}}}
\;e^{-\frac{\Phi}{4}}
\rho_2\,K_aK_b\;\delta(\chi-\chi_0),
&
(a,b=0,\theta,\phi, p)
\vspace{12pt}
\\
\displaystyle{\frac{\mu_2}{V^{\ast}}}
e^{\frac{3\Phi}{4}}
\rho_2\,G_{ab}\;\delta(\chi-\chi_0),
&
(a,b=m)
\vspace{12pt}
\\
0,
&
({\rm otherwise})
\end{array}
\right.
\nonumber
\end{eqnarray}
\vspace{-18pt}
\begin{eqnarray}
\end{eqnarray}

It is interesting to note that the energy-momentum tensor
has several particular features. Firstly, the dominant
energy condition is violated near the enhan\c{c}on
radius $r=r_e$ because
$T^{D2}_{00}+T^{D6}_{00}$ is proportional to
$(V/Z_6-V_*/Z_2)\sim 0$
and hence $|T^{D2}_{{\bar{0}\bar{0}}}
+T^{D6}_{\bar{0}\bar{0}}|\ll
|T^{D2}_{\bar{p}\bar{q}}+T^{D6}_{\bar{p}\bar{q}}|$,
where $\bar{p}$, $\bar{q}$ implies
an orthogonal basis. Secondly, the
energy-momentum tensor looks like dust in the three-dimensional
subspace $(x^0,\theta,\phi)$, while it looks like a
varying cosmological constant in the other subspace $(x^0,x^4,x^5)$,
due to along which directions D6-D2 branes are aligned.

\section{Israel's Junction condition and
enahn\c{c}on geometry }
\label{sec.IV}

If the shell sits on a radius $r=r_0~(>r_e)$,
timelike hypersurface $\chi=\chi_0$ corresponds to $r=r_0$
one. On the hypersurface, Israel's junction
equations~\cite{Israel} are
\begin{eqnarray}
\label{eq-junction}
[\Pi_{ab}]_\pm :=[K_{ab}-q_{ab}\,{K^c}_c]_\pm
=-{\kappa_{10}}^2
(S^{D6}_{ab}+S^{D2}_{ab}),
\end{eqnarray}
where $S^{D6}_{ab}$ and $S^{D2}_{ab}$ are
functions in front of $\delta$-function
of ${T}^{D6}_{ab}$ and ${T}^{D2}_{ab}$, respectively.
Using the unit normal vector $n_M=(d\chi)_M$, the second
fundamental form $K_{ab}$ is defined by
$K_{ab} := n_{a;b}$. $[f]_\pm$ simply denotes
$f_+-f_-$, where
$f_+$ and $f_-$ are quantities evaluated on the
outside and inside of the $r=r_0$ timelike hypersurface,
respectively.

According to the enhan\c{c}on mechanism, the inside of
the shell
is replaced with a flat geometry. Reminding that
$G_{MN}=e^{-\frac{\Phi}{2}}\hat{G}_{MN}$, we obtain
\begin{eqnarray}
\label{eq-normal}
d\chi={Z_2}^{\frac{3}{16}}{Z_6}^{\frac{7}{16}}dr.
\end{eqnarray}
Thus, we can easily calculate all components of
$K_{ab}$ at any $r=r_0$ :
\begin{eqnarray}
\label{extrinsic curv_in-uu}
K_{00-}=0,
\end{eqnarray}
\begin{eqnarray}
\label{extrinsic curv_in-tt}
K_{\theta\theta-}=re^{-\frac14 \phi}(Z_2Z_6)^{\frac14},
\end{eqnarray}
\begin{eqnarray}
\label{extrinsic curv_in-mn}
K_{mn-}=0,
\end{eqnarray}
\begin{eqnarray}
\label{extrinsic curv_in-pq}
K_{pq-}=0,
\end{eqnarray}
and
\begin{eqnarray}
\label{extrinsic curv_out-uu}
K_{00+}=\frac{1}{16}e^{-\frac14 \phi}(Z_2Z_6)^{-\frac34}
\left(\frac{5Z_2'}{Z_2}+\frac{Z_6'}{Z_6}\right),
\end{eqnarray}
\begin{eqnarray}
\label{extrinsic curv_out-tt}
K_{\theta\theta+}=\frac{r^2}{16}e^{-\frac14 \phi}(Z_2Z_6)^{\frac14}
\left(\frac{3Z_2'}{Z_2}+\frac{7Z_6'}{Z_6}+\frac{16}{r}\right),
\end{eqnarray}
\begin{eqnarray}
\label{extrinsic curv_out-mn}
K_{mn+}=-\frac{1}{16}e^{-\frac14 \phi}(Z_2Z_6)^{-\frac34}
\left(\frac{5Z_2'}{Z_2}+\frac{Z_6'}{Z_6}\right)\delta_{mn},
\end{eqnarray}
\begin{eqnarray}
\label{extrinsic curv_out-pq}
K_{pq+}=\frac{1}{16}e^{-\frac14
\phi}V^{\frac12}Z_2^{\frac14}Z_6^{-\frac34}
\left(\frac{3Z_2'}{Z_2}-\frac{Z_6'}{Z_6}\right)G^{(K3)}_{pq}.
\end{eqnarray}

We can check that
Eq.~(\ref{eq-junction}) is satisfied at
any radius $r=r_0$ with the help of
Eqs.~(\ref{conservation1}) and (\ref{conservation4}).
So, we obtain a whole static solution
without any repulson singularity, taking the
self-gravity of the shell into account. This is
one of the main results in this paper\cite{com1}.

Next, let us consider the whole energy-momentum
tensor $T_{MN}$ coming from both supergravity and
the shell parts and derive the energy conservation
on the shell.
In terms of the unit timelike vector $u^M$ on the shell,
the energy conservation is written by $u^M\nabla^N T_{MN}=0$.
To derive the conservation equation effectively
on the shell, we shall put a step function
$\theta(\chi-\chi_0)$ in front of the
matter Lagrangian density of the supergravity part
in Eq.~(\ref{E-action})\footnote{We thank to Kouji
Nakamura for suggesting this.}, i.e.,
$\int\,d^{10}x\,\sqrt{-G}{\cal L}_m
\to \int\,d^{10}x\,\sqrt{-G}\,
\theta(\chi-\chi_0){\cal L}_m$, where
${\cal L}_m=-(\nabla \Phi)^2/2
-e^{\Phi/2}{F_4}^2/2\cdot 4!-
e^{-3\Phi/2}{F_8}^2/2\cdot 8!$.
Then, $T_{MN}$ is derived by the action~(\ref{E-action})
as follows:
\begin{eqnarray}
\label{W-energy-momentum}
T_{MN}&=&\delta(\chi-\chi_0)\,
{{S}^{D6 \mbox{-} D2}}_{MN}
\nonumber \\
&& \;\;
+\frac{\theta(\chi-\chi_0)}{{\kappa_{10}}^2}\,
({S^{\rm RR}}_{MN}+
{S^{\rm dilaton}}_{MN}),
\end{eqnarray}
where
\begin{eqnarray}
\label{E-energy-momentum}
{S^{\rm RR}}_{MN}&=&\frac{e^{\frac{\Phi}{2}}}{4!}
\,\Bigl(4{F_{(4)}}_{MLJK}{{F_{(4)}}_N}^{LJK}
-\frac{1}{2}G_{MN}{F_{(4)}}^2\Bigl)
\nonumber \\
&& +\frac{e^{-\frac{3\Phi}{2}}}{8!}
\,\Bigl(8{F_{(8)}}_{MM_2M_3...M_8}
{{F_{(8)}}_N}^{M_2M_3...M_8}
\nonumber \\
&& \;\;\;\;\;\;\;\;
-\frac{1}{2}G_{MN}{F_{(8)}}^2 \Bigr),
\end{eqnarray}
\begin{eqnarray}
\label{P-energy-momentum}
{S^{\rm dilaton}}_{MN}=
\nabla_M\Phi\nabla_N\Phi-\frac{1}{2}G_{MN}
(\nabla\Phi)^2
\end{eqnarray}
and ${{S}^{D6 \mbox{-} D2}}_{MN} :=
{{S}^{D6}}_{MN}+{{S}^{D2}}_{MN}$.
Note that ${{S}^{D6 \mbox{-} D2}}_{\chi a}=
{{S}^{D6 \mbox{-} D2}}_{\chi\chi}=0$.
Because we introduced a step function, all components of
${S^{\rm RR}}_{MN}$ and ${S^{\rm dilaton}}_{MN}$ are
smooth functions of $\chi$ and they can be calculated
on the outside of the shell.
Then, $u^M\nabla^N T_{MN}=0$ is written by
the Gaussian normal coordinates~(\ref{E-gaussian}) as
\begin{eqnarray}
\label{D-energy-momentum1}
0 &=&u^M\nabla^N{T}_{MN}
\nonumber \\
&=&u^M\biggl[D^N {S^{{D6 \mbox{-} D2}}}_{MN}
\left.+\frac{1}{{\kappa_{10}}^2}
\bigl({S^{\rm RR}}_{M\chi}+{S^{\rm dilaton}}_{M\chi}\bigr)
\right]
\nonumber \\
&& \;\;\;\times\delta(\chi-\chi_0)
+\mbox{(regular functions of}
\,\,\chi\,),
\end{eqnarray}
where $D$ denotes covariant derivative with respect to the metric
$q_{MN}=G_{MN}-n_Mn_N$.
Here, a regular function $f(x)$ means that
$f(x)$ is integrable and
$\int^{x_0+\epsilon}_{x_0-\epsilon}f(x)dx\sim O(\epsilon)$.
Integrating Eq.~(\ref{D-energy-momentum1}) from
$\chi=\chi_0-\epsilon$ to
$\chi=\chi_0+\epsilon$, we obtain the following equations:
\begin{eqnarray}
\label{D-energy-momentum3}
&&{\kappa_{10}}^2 u^MD^N {S^{D6\mbox{-}D2}}_{MN}
\nonumber \\
&& \;\;\;\;\;\;\;
+({S^{\rm RR}}_{MN}+{S^{\rm dilaton}}_{MN})u^M n^N=0.
\end{eqnarray}
As easily seen, this equation is always satisfied when the shell is static.

Finally, let us consider the dilaton
equation.
Varying the action (\ref{E-action}) with respect to $\Phi$,
we get the
dilaton field equation
\begin{eqnarray}
\label{dilaton}
\nabla^2\Phi&=&
\frac{e^{\frac{1}{2}\Phi}}{4\cdot4!}\,{F_{(4)}}^2
-\frac{3e^{-\frac{3}{2}\Phi}}{4\cdot8!}\,{F_{(8)}}^2
\nonumber \\
&&-2{\kappa_{10}}^2\frac{\delta}{\delta \Phi}
\sum_{i=1}S^{D6\mbox{-}D2}_{(i)}.
\end{eqnarray}
In terms of Gaussian normal coordinates~(\ref{E-gaussian}), the
left hand side of Eq.~(\ref{dilaton}) is expressed as
\begin{equation}
\label{dilatonG}
\nabla^2 \Phi=\frac{1}{\sqrt{-G}}
\left[\frac{\partial}{\partial\chi}(\sqrt{-G}\Phi_{,\chi})
+\frac{\partial}{\partial \zeta^0}(\sqrt{-G}q^{00}\Phi_{,0})
\right].
\end{equation}
Since the second term is a regular function of $\chi$,
integrating Eq.~(\ref{dilaton}) from $\chi=\chi_0-\epsilon$ to
$\chi=\chi_0+\epsilon$, we get a junction condition for the dilaton
field such as
\begin{eqnarray}
\label{junctionD}
[\Phi_{,\chi}]_\pm=\frac{{\kappa_{10}}^2 \mu_6}{2}\,
\Bigl(V_*\,e^{-\frac{\Phi}{4}}\rho_2+3\,
e^{\frac{3\Phi}{4}}\rho_6\Bigr).
\end{eqnarray}

In the following sections, we will consider the
perturbation of the solution.

\section{Stability analysis  of the D6-D2 brane shell}
\label{sec.V}

To avoid complication, let us focus on the following
perturbed metric in Einstein frame.
\begin{eqnarray}
\label{pM}
ds^2&=&e^{-\frac{\Psi_\pm}{2}}\bigl({Z_{2\pm}}^{-\frac{1}{2}}
{Z_{6\pm}}^{-\frac{1}{2}}
\eta_{\mu\nu} dx^\mu dx^\nu
\nonumber \\
&& \;\;\;\;\;\;\;\;\;\;
+{Z_{2\pm}}^{\frac{1}{2}}{Z_{6\pm}}^\frac{1}{2} dx^i dx^i
\nonumber \\
&&\;\;\;\;\;\;\;\;\;\;
+V^{\frac{1}{2}} {Z_{2\pm}}^\frac{1}{2}
{Z_{6\pm}}^{-\frac{1}{2}}
G^{K3}_{pq}dx^p dx^q\bigr),
\end{eqnarray}
where
\begin{eqnarray}
\Psi_\pm(t,r)&=&\bar{\Phi}_\pm(r)+\delta\Psi_\pm(t,r),
\nonumber \\
Z_{2\pm}(t,r)&=&\bar{Z}_{2\pm}(r)
\bigl[1+\delta Z_{2\pm}(t,r)\bigr],
\nonumber \\
Z_{6\pm}(t,r)&=&\bar{Z}_{6\pm}(r)
\bigl[1+\delta Z_{6\pm}(t,r)\bigr],
\end{eqnarray}
where $\pm$ denotes outside and inside of the shell.
$\bar{f}$ denotes an unperturbed background field $f$.
The dilaton field $\Phi_\pm$ and Ramond-Ramond fields
$C_{(3)\pm}$ and $C_{(7)\pm}$ are also written by
\begin{eqnarray}
\label{pmatter}
\Phi_\pm(t,r)&=&\bar{\Phi}_\pm(r)+\delta\Phi_\pm(t,r),
\nonumber \\
C_{3\pm}(t,r)&=&\bar{C}_{3\pm}(r)
+\delta C_{3\pm}(t,r),
\nonumber \\
C_{7\pm}(t,r)&=&\bar{C}_{7\pm}(r)
+\delta C_{7\pm}(t,r).
\end{eqnarray}
Here the fields $C_3$ and $C_7$ are components of the
Ramond-Ramond
fields $C_{(3)}$ and $C_{(7)}$ respectively:
\begin{eqnarray}
\label{R-R fields}
C_{(3)}&=&C_{3}\;dt\wedge dx^4\wedge dx^5.
\\
C_{(7)}&=&VC_{7}\;dt\wedge dx^4\wedge dx^5\wedge
dV_{(K3)}.
\end{eqnarray}
The above unperturbed fields are defined as
\begin{eqnarray}
\label{Unperturbed Z}
\bar{Z}_2(r) := \left\{
\begin{array}{ll}
\bar{Z}_{2-}(r)=1 + r_2/r_0  \\
\bar{Z}_{2+}(r)=1 + r_2/r
\end{array}
\right. \\
\bar{Z}_6(r) := \left\{
\begin{array}{ll}
\bar{Z}_{6-}(r)=1 + r_6/r_0  \\
\bar{Z}_{6+}(r)=1 + r_6/r
\end{array}
\right.
\end{eqnarray}
and $\bar{\Phi}_\pm(r)$, $\bar{C}_{3\pm}(r)$ and
$\bar{C}_{7\pm}(r)$ are given by
$\bar{Z}_{2\pm}(r)$ and $\bar{Z}_{6\pm}(r)$.

Assume that the shell sits on $r=r_0$ before it is
perturbed.
Since the inside of the geometry is just flat, all functions
$\bar{f}_-$ are constants determined by the continuity
of the outside and inside of the metric, i.e.,
$\bar{f}_-=\bar{f}_+(r=r_0)$.

\subsection{The perturbation of the
flat geometry inside the metric}
\label{sec.V-i}

The perturbed energy-momentum tensor $\delta{T}_{MN-}$
does not include any linear terms of $\delta\Phi_-$,
$\delta C_{3-}$ or $\delta C_{7-}$ because the
background energy-momentum tensor $T_{MN-}$ is zero.
Hence, the perturbed Einstein equation is simply
$\delta R_{MN-}=0$.
Defining derivative operator
$\bar{\nabla}^2_- :=\partial_t \partial^t+
\partial_i\partial^i$,
the equations are written by

\begin{eqnarray}
\label{EinM-1}
&&\bar{\nabla}^2_-(\delta\Psi_-+\delta Z_{2-}+\delta Z_{6-})
\nonumber \\
&&\;\;\;\;\;\;\;\;\;\;
-\partial_t^2\,(8\,\delta\Psi_--6
\,\delta Z_{2-}+2
\,\delta Z_{6-})=0,
\end{eqnarray}
\begin{eqnarray}
\label{EinM-2}
\partial_t\partial_i\,(2\,\delta\Psi_--\delta Z_{2-}+
\delta Z_{6-})=0,
\end{eqnarray}
\begin{eqnarray}
\label{EinM-3}
\bar\nabla^2_-\,(\delta\Psi_-+\delta Z_{2-}+\delta Z_{6-})=0,
\end{eqnarray}
\begin{eqnarray}
\label{EinM-4}
&& \bar\nabla^2_-\,(\delta\Psi_--\delta Z_{2-}-\delta Z_{6-})
\;\delta_{ij}
\nonumber \\
&&\;\;\;\;\;\;\;\;\;\;
+\partial_i\partial_j
\,(8\,\delta\Psi_--2\,\delta Z_{2-}+6\,\delta Z_{6-})=0,
\end{eqnarray}
\begin{eqnarray}
\label{EinM-5}
\bar{\nabla}^2_-(\delta\Psi_--\delta Z_{2-}+\delta Z_{6-})=0.
\end{eqnarray}

If we take
\begin{eqnarray}
\delta\Psi_{\pm} & =&\xi_{\pm}(r)\,e^{i\omega t},
\label{decouple1}\\
\delta Z_{2\pm}& =&\zeta_{2\pm}(r)\,e^{i\omega t}, \\
\delta Z_{6\pm}& =&\zeta_{6\pm}(r)\,e^{i\omega t}, \\
\delta \Phi_{\pm}& =&\eta_{\pm}(r)\,e^{i\omega t},
\label{decouple4}
\end{eqnarray}
Eq.~(\ref{EinM-2}) indicates
\begin{eqnarray}
\label{EinM-6}
\xi_-=\frac{1}{2}(\zeta_{2-}-\zeta_{6-}).
\end{eqnarray}
Substituting this into Eqs.~(\ref{EinM-1})-(\ref{EinM-5}),
we obtain the following equations
\begin{eqnarray}
\label{EinM-7}
&\bar\nabla^2_- \delta\Psi_{-}=0,&
\end{eqnarray}
\begin{eqnarray}
\label{relation-}
&\xi_-=\zeta_{2-}=-\zeta_{6-}.&
\end{eqnarray}
The perturbed field equation of the dilaton is same as Eq.~(\ref{EinM-7}):
\begin{eqnarray}
\label{EinM-8}
\bar\nabla^2_- \delta\Phi_{-}=0.
\end{eqnarray}

Imposing a regular boundary condition at $r=0$,
the solution of Eqs.~(\ref{EinM-7}) and (\ref{EinM-8}) are
\begin{eqnarray}
\label{EinM-S}
\xi_-(r)=A_{\Psi}\,\frac{\sin(\Omega r)}{\Omega r},
\end{eqnarray}
\begin{eqnarray}
\label{sol-8}
\eta_-(r)=A_{\Phi}\,\frac{\sin(\Omega r)}{\Omega r},
\end{eqnarray}
where $\Omega=\sqrt{\bar{Z}_{2-}\bar{Z}_{6-}}\,\omega$,
$A_{\Psi}$ and $A_{\Phi}$ are integral constants.
Note that when the shell is absent, the frequency
$\omega$ must be real. Otherwise, i.e.,
when $\omega$ is imaginary, the eigenfunctions
diverge at infinity.
However, when the shell exists at some radius,
the gradient of eigenfunctions becomes discontinuous
at the shell and there is a possibility that the
exponential growth of the eigenfunction may be
changed by the exponentially decaying function
at the shell. In this sense, the stability of the
shell is not trivial.

\subsection{The perturbation of the junction
conditions on the shell}
\label{sec.V-ii}

Suppose that the orbit of the shell is
$r=r_0+\delta R(t)$ in
the metric~(\ref{pM}).
In this and next subsections, we mainly discuss the values
around the shell. Hence all the variables are
evaluated on the shell if we do not mention in these
subsections. We also drop the index ${\pm}$ of the
continuous variables on the shell.
Then, the (linear order) continuity of
the metric and dilaton field yields the following
equations,
\begin{eqnarray}
\label{junc-1}
\delta\Phi_+&=&\delta\Phi_-
-{\bar{\Phi}_+}'\,\delta R,
\nonumber \\
\delta Z_{2+}&=&\delta Z_{2-}-
\frac{\bar{Z}_{2+}'}{\bar{Z}_{2}}
\,\delta R,
\nonumber \\
\delta Z_{6+}&=&\delta Z_{6-}-
\frac{\bar{Z}_{6+}'}{\bar{Z}_{6}}
\,\delta R,
\end{eqnarray}
where a prime denotes differential with respect to $r$.

Firstly, let us solve Eq.~(\ref{D-energy-momentum3}) and show that
the dilaton field $\Phi$ is constant along the shell, i.e.,
$d\Phi/d\zeta^0=0$.
The components of the vectors
$u^M=\bar{u}^M+\delta u^M$ and $n^M=\bar{n}^M+\delta n^M$
are expressed as
\begin{eqnarray}
\label{vector}
\bar{n}^M&=&(\bar{n}^t,\bar{n}^r,0,\cdots,0)
\nonumber \\
&=&
\bigl(0,{\bar{Z}_2}^{-\frac{3}{16}}
{\bar{Z}_6}^{-\frac{7}{16}},0,\cdots,0\bigr)
\nonumber \\
\delta{n}^M&=&
\Bigl({\bar{Z}_2}^{\frac{13}{16}}{\bar{Z}_6}^
{\frac{9}{16}}\delta\dot{R},
\nonumber \\
&& \;\;\;\;
\frac{1}{4}{\bar{Z}_2}^{-\frac{3}{16}}
{\bar{Z}_6}^{-\frac{7}{16}}
(\delta\Psi-\delta Z_2-\delta Z_6 )
,0,\cdots,0 \Bigr)
\nonumber \\
\bar{u}^M&=&\bigl({\bar{Z}_2}^{\frac{5}{16}}{\bar{Z}_6}^
{\frac{1}{16}},0,0,\cdots,0\bigr),
\nonumber \\
\delta{u}^M&=&
\Bigl(\frac{1}{4}{\bar{Z}_2}^{\frac{5}{16}}
{\bar{Z}_6}^{\frac{1}{16}}
(\delta\Psi+\delta Z_2+\delta Z_6),
\nonumber \\
&& \;\;\;\;
{{\bar{Z}}_2}^{\frac{5}{16}}{\bar{Z}_6}^{\frac{1}{16}}
\delta\dot{R}
,0,\cdots,0 \Bigr),
\end{eqnarray}
where dot means derivative with respect to $t$.
After straightforward but long and tedious calculation, we get
${\delta(S^{\rm dilaton}}_{MN}u^M n^N)$
and $\delta({S^{\rm RR}}_{MN}u^M n^N)$
as
\begin{eqnarray}
\label{D-energy-momentum4}
&& {\delta ({S^{\rm dilaton}}_{MN}u^M n^N)|}_{\chi=\chi_0}
\nonumber \\
&&\;\;\;\;\;\;\;\;\;=
\frac{1}{8}{\bar{Z}_{2}}^{\frac{1}{8}}
{\bar{Z}_{6}}^{-\frac{3}{8}}
\left(\frac{\bar{Z}_{2+}'}{\bar{Z}_{2}}
-\frac{3\bar{Z}_{6+}'}{\bar{Z}_{6}}\right)
\nonumber \\
&& \;\;\;\;\;\;\;\;\;\;\;\;\;\;\;\;\;\;
\times\left\{\delta\dot{\Phi}_++\frac{1}{4}
\delta\dot{R}\left(\frac{\bar{Z}_{2+}'}{\bar{Z}_{2}}
-\frac{3\bar{Z}_{6+}'}{\bar{Z}_{6}}\right)\right\}
\Biggl|_{r=r_0}
\nonumber \\
&& \;\;\;\;\;\;\;\;\;=
\frac{1}{8}{\bar{Z}_{2}}^{\frac{1}{8}}
{\bar{Z}_{6}}^{-\frac{3}{8}}
\left(\frac{\bar{Z}_{2+}'}{\bar{Z}_{2}}
-\frac{3\bar{Z}_{6+}'}{\bar{Z}_{6}}\right)
\delta\dot{\Phi}_-\Biggl|_{r=r_0},
\end{eqnarray}
\begin{eqnarray}
{\delta ({S^{\rm RR}}_{MN}u^M n^N)|}_{\chi=\chi_0}&=&0,
\end{eqnarray}
where we used Eq.~(\ref{junc-1}) to derive the second equation.

The first term in Eq.~(\ref{D-energy-momentum3}) is rewritten by
\begin{eqnarray}
\label{D-energy-momentum5}
&& u^M\nabla^N {S^{D6\mbox{-}D2}}_{MN}
\nonumber \\
&& \;\;\;\;\;
=\Bigl[\mu_6 u^M\partial_M W
-\frac{\mu_2}{2}{G^{(K3)}}^{pq}u^M
\partial_M G_{pq}\,
\rho_2\,e^{-\frac{\Phi}{4}}
\nonumber \\
&& \;\;\;\;\;\;\;\;\;
+\frac{\mu_6}{2}(G^{\theta\theta}u^M\partial_M
G_{\theta\theta}
+G^{\phi\phi}u^M\partial_M G_{\phi\phi})W\Bigr]
\Bigl|_{\chi=\chi_0},
\nonumber \\
\end{eqnarray}
where
\begin{eqnarray}
W&=&\rho_6\,e^{\frac{3\Phi}{4}}-
V_*\rho_2\,e^{-\frac{\Phi}{4}}.
\end{eqnarray}
This quantity can be expanded as
${\overline{u^M\nabla^N {S^{D6\mbox{-}D2}}_{MN}}}|_{r=r_0}
+\delta(u^M\nabla^N {S^{D6\mbox{-}D2}}_{MN})$
in terms of both the metrics outside and inside the shell. Therefore,
it is convenient to expand Eq.~(\ref{D-energy-momentum5}) in terms of
the metric inside the shell. As a result, we get
\begin{eqnarray}
\label{D-energy-momentum6}
&&{\kappa_{10}}^2\delta\bigl(u^M\nabla^N
{S^{D6\mbox{-}D2}}_{MN}\bigr)
\nonumber \\
&& \;\;\;\;\;\;\;\;=
-\frac{1}{8r_0^2}\,
{\bar{Z}_2}^{\frac{1}{8}}{\bar{Z}_6}^{-\frac{3}{8}}
\left(\frac{r_2}{\bar{Z}_2}+\frac{r_6}{\bar{Z}_6}
\right)\delta\dot{\Phi}_-
\end{eqnarray}
with help of Eqs.~(\ref{relation-}) and (\ref{vector}).
Finally, substituting Eqs.~(\ref{D-energy-momentum4}) and
(\ref{D-energy-momentum6}) into
Eq.~(\ref{D-energy-momentum3}),
we get the following simple equation
\begin{eqnarray}
\label{D-energy-momentum7}
\delta\dot{\Phi}_-=0,
\end{eqnarray}
which means that the dilaton field is constant
along the shell in the first order of the
perturbation. It should be noted, however, that
it does not necessarily imply
$\delta \Phi_-{(r)}=0$.
For example, the general solution of the perturbed
dilaton field equation is (\ref{sol-8}).
When the frequency
$\Omega$ (or $\omega$) is
real, the condition (\ref{D-energy-momentum7})
means that the $\delta \Phi_-(r)$ becomes node
at $r=r_0$. The frequency of the perturbation
becomes discrete to satisfy this condition when the
perturbation of the dilaton field has nontrivial configuration.
Of course, there may exist a solution with $A_{\Phi}=0$.
In this case, the frequency shows continuous spectrum.
On the other hand, when $\omega$ is imaginary,
the eigenfunction grows exponentially with $r$, and
it cannot become zero unless $A_{\Phi}=0$. Hence
$\delta \Phi_{-} \equiv 0$ for the imaginary
frequency, i.e., unstable modes.

Now, let us focus on the perturbation of the junction
equations~(\ref{eq-junction}). It is noteworthy that
the right hand side does not include any $\delta\Phi_-$
terms, provided that we calculate it by using the inside
metric~(\ref{pM}),
as shown above. So, all non-vanishing equations are:
\begin{eqnarray}
\label{00,mn-pi}
&&\delta Z'_{2+}-\delta Z'_{6+}
-2\delta \Psi'_+
=\frac{1}{8r_0^2}
\left(\frac{3\,r_2}{\bar{Z}_2}-\frac{r_6}{\bar{Z}_6}
\right)
\delta Z_{2-}
\nonumber \\
&&\;\;\;\;\;\;\;\;\;\;\;\;\;\;\;\;
+\frac{1}{2r_0^4}
\left(\frac{{r_2}^2}{{\bar{Z}_2}^2}
+\frac{{r_6}^2}{{\bar{Z}_6}^2} \right)
{\delta R},
\end{eqnarray}
\begin{eqnarray}
\label{pq-pi}
&&\delta Z'_{2+}-2\delta Z'_{6+}
-4\delta \Psi'_{+}
\nonumber \\
&&\;\;\;\;\;\;\;\;\;\;\;\;\;\;\;\;
=-\frac{1}{4r_0^2}
\frac{r_6}{\bar{Z}_6}
{\delta Z_{2-}}
-\delta {Z_{2-}'}
+\frac{1}{r_0^4}\frac{{r_6}^2}{{\bar{Z}_6}^2}
{\delta R},
\end{eqnarray}
\begin{eqnarray}
\label{theta-pi}
\delta Z'_{2+}-
3\delta Z'_{6+}-4\delta \Psi'_{+}
=-\frac{1}{8r_0^3}
\left(\frac{3\,r_2}{\bar{Z}_2}+
\frac{7\,r_6}{\bar{Z}_6}\right)
\delta R.
\end{eqnarray}
By solving these equations with respect
to $\delta {\Psi}'_{+}$, $
\delta {Z'_{2+}}$,
and $\delta {Z'_{6+}}$, we obtain
\begin{eqnarray}
\label{deltaZ_2}
\delta Z'_{2+}-\delta Z'_{2-}
=\frac{3}{4{r_0}^2}
\frac{r_2}{\bar{Z}_2}
\;{\delta Z^-_2}
+\frac{1}{{r_0}^4}
\frac{{r_2}^2}{{\bar{Z}_2}^2}
\;{\delta R},
\end{eqnarray}
\begin{eqnarray}
\label{deltaZ_6}
&&\delta Z'_{6+}-\delta Z'_{6-}
=
-\frac{1}{4{r_0}^2}
\frac{r_6}{\bar{Z}_6}
{\delta Z_{2-}}
\nonumber \\
&& \;\;\;\;\;\;\;\;\;\;
+\frac{1}{8{r_0}^3}\left(
\frac{3r_2}{\bar{Z}_2}+\frac{7r_6}{\bar{Z}_6}
+\frac{8}{r_0}
\frac{{r_6}^2}{{\bar{Z}_6}^2}\right)
{\delta R},
\end{eqnarray}
\begin{eqnarray}
\label{delta-Psi}
&& \delta \Psi'_{+}-\delta\Psi'_{-}
=
\frac{3}{16{r_0}^2}
\left(\frac{r_2}{\bar{Z}_2}+
\frac{r_6}{4\bar{Z}_6}\right)
{\delta Z_{2-}}
\nonumber \\
&& \;\;\;\;\;\;\;
-\frac{1}{16{r_0}^3}\left[
\frac{3r_2}{\bar{Z}_2}+\frac{7r_6}{\bar{Z}_6}
-\frac{4}{{r_0}}
\left(\frac{{r_2}^2}{{\bar{Z}_2}^2}
-\frac{3{r_6}^2}{{\bar{Z}_6}^2} \right)\right]
{\delta R}.
\end{eqnarray}
The left hand side in
Eqs.~(\ref{deltaZ_2})-(\ref{delta-Psi}) are
interpreted as a difference between inside and outside
quantities. So, if the number densities $\rho_2$ and
$\rho_6$
of D2 and D6 branes on the shell approach zero,
the difference also approaches zero.

Similarly, we can obtain $\delta\Phi'_+$ from the junction
condition of the dilaton field
Eq.~(\ref{junctionD})
as follows:
\begin{eqnarray}
\label{delta-Phi}
\delta \Phi'_{+} -\delta \Phi'_{-}&=&-\frac{3}{16{r_0}^2}
\left(\frac{r_2}{\bar{Z}_2}+\frac{r_6}{\bar{Z}_6}\right)
{\delta Z_{2-}}
\nonumber \\
&& \;\;\;\;\;\;\;\;\;\;
+\frac{1}{4{r_0}^4}
\left(\frac{{r_2}^2}{{\bar{Z}_2}^2}
-\frac{3{r_6}^2}{{\bar{Z}_6}^2}\right)
{\delta R}.
\end{eqnarray}
We also have the following equations from the RR
field equations using Eq.~(\ref{junc-1}):
\begin{eqnarray}
\label{Junction C7}
\delta C_{7+}^{\prime} =
\frac{r_6}{2gr_0^2\bar{Z}_6^2}\Bigl(
\delta \Psi_{-}+3\delta \Phi_{-}\Bigr)
-\frac{2r_{6}^{2}}{gr_0^4\bar{Z}_6^3}\delta R,
\end{eqnarray}
\begin{eqnarray}
\label{Junction C3}
\delta C_{3+}^{\prime} =
-\frac{r_2}{2gr_0^2\bar{Z}_2^2}\Bigl(3
\delta \Psi_{-}+\delta \Phi_{-}\Bigr)
-\frac{2r_2^2}{gr_0^4\bar{Z}_2^3}\delta R.
\end{eqnarray}

\subsection{Equation of motion of the shell}
\label{sec.V-iii}

When the locus of the shell made of the wrapped D6-branes
shifts from $r=r_0$ to $r=R(t)=r_0 + \delta R(t)$,
where $|\delta R(t)|\ll r_0$, the effective action of the
shell up to the second order of $\delta \dot{R}(t)$
is given by
\begin{eqnarray}
\label{Perturb-action}
S^{D6\mbox{-}D2}&=&-N\int d^3\xi\,
e^{\frac{3}{4}(\Phi-\Psi)}\,
{Z_2}^{\frac{1}{4}} {Z_{6}}^{-\frac{3}{4}}
\nonumber \\
&& \;\;\;\;
\left(\frac{\mu_6V}{Z_6}\, e^{-\Psi}-\frac{\mu_2}{Z_2}\,
e^{-\Phi}\right)
\Bigl[1-\frac{1}{2}Z_2Z_6(\delta\dot{R})^2\Bigl]
\nonumber \\
&&+N\int d^3\xi\,\Bigl(\mu_6VC_7-\mu_2C_3\Bigl).
\end{eqnarray}
{}From the action (\ref{Perturb-action}),
we have the following
equation of motion of the shell:
\begin{eqnarray}
\label{Eq of d2d6}
&&\Bigl[\mu_6V\bar{Z}_2(R)-\mu_2\bar{Z}_6(R)\Bigl]
\delta \ddot{R}
\nonumber \\
&&\;\;\;\;
=-\frac{\mu_6V}{4\bar{Z}_6(R)}
\biggl[-\frac{\bar{Z}_6^{\prime}(R)}{\bar{Z}_6(R)}
(3\delta\Phi-7\delta\Psi+\delta Z_2-7\delta Z_6)
\nonumber \\
&& \;\;\;\;\;\;\;\;\;\;\;\;
+(3\delta\Phi-7\delta\Psi+
\delta Z_2-7\delta Z_6)^{\prime}\biggl]
\nonumber \\
&&\;\;\;\;\;\;\;
\,\,-\frac{\mu_2}{4\bar{Z}_2(R)}
\biggl[-\frac{\bar{Z}_2^{\prime}(R)}{\bar{Z}_2(R)}
(\delta\Phi+3\delta\Psi+3\delta Z_2+3\delta Z_6)
\nonumber \\
&& \;\;\;\;\;\;\;\;\;\;\;\;
+(\delta\Phi+3\delta\Psi+
3\delta Z_2+3\delta Z_6)^{\prime}\biggl]
\nonumber \\
&&\;\;\;\;\;\;\;\;
+g\mu_6V\delta C_7^{\prime}
- g\mu_2\delta C_3^{\prime},
\end{eqnarray}
where the primes denote differentials with respect to
$R$. $\delta C_3$ and $\delta C_7$  are the
perturbed parts of
the Ramond-Ramond fields respectively.

The fields in Eq.~(\ref{Eq of d2d6}) are discontinuous
at the shell $r=R(t)$, so that the equation needs to be
averaged near the shell.
Recalling that the average of $A(R)$ at the shell,
$\{A\}$, is
defined by $\{A\}=\frac{1}{2}(A_{-}+A_{+})$,
the equation of motion of the shell up to the first
order of the perturbation can be written as
\begin{eqnarray}
\label{Averaged Eq-of-Mot}
&&(\mu_6V\bar{Z}_2-\mu_2\bar{Z}_6)
\delta \ddot{R}
\nonumber \\
&&\;\;\;\;\;\;
=-\frac{\mu_6V}{8\bar{Z}_6}
\bigl(3\delta\Phi_--7\delta\Psi_-
+\delta Z_{2-}-7\delta Z_{6-}\bigr)^{\prime}
\nonumber \\
&&\;\;\;\;\;\;\;\;\;\;\;\;
-\frac{\mu_2}{8\bar{Z}_2}
\bigl(\delta\Phi_-+3\delta\Psi_-
+3\delta Z_{2-}+3\delta Z_{6-}\bigr)^{\prime}
\nonumber \\
&&\;\;\;\;\;\;\;\;\;\;\;\;
-\frac{\mu_6V}{8\bar{Z}_6}
\biggl[-\frac{\bar{Z}_6^{\prime}}{\bar{Z}_6}
\bigl(3\delta\Phi_{+}-7\delta\Psi_{+}
+\delta Z_{2+}-7\delta Z_{6+}\bigr)
\nonumber \\
&&\;\;\;\;\;\;\;\;\;\;\;\;\;\;\;\;\;\;
+\bigl(3\delta\Phi_+-7\delta\Psi_++
\delta Z_{2+}-7\delta Z_{6+}\bigr)^{\prime}\biggl]
\nonumber \\
&&\;\;\;\;\;\;\;\;\;\;\;\;
-\,\frac{\mu_2}{8\bar{Z}_2}
\biggl[-\frac{\bar{Z}_2^{\prime}}{\bar{Z}_2}
\bigl(\delta\Phi_{+}+3\delta\Psi_{+}+
3\delta Z_{2+}+3\delta Z_{6+}\bigr)
\nonumber \\
&&\;\;\;\;\;\;\;\;\;\;\;\;\;\;\;\;\;\;
+\bigl(\delta\Phi_++3\delta\Psi_++
3\delta Z_{2+}+3\delta Z_{6+}\bigr)^{\prime}\biggl]
\nonumber \\
&&\;\;\;\;\;\;\;\;\;\;\;\;
+\,\frac{g}{2}\mu_6V\bigl(\delta C_{7-}
+\delta C_{7+}\bigr)^{\prime}-\frac{g}{2}\mu_2
\bigl(\delta C_{3-}
+\delta C_{3+}\bigr)^{\prime},
\nonumber \\
\end{eqnarray}
where we can take
$\bar{Z}_2^{\prime} =-{r_2}/{{r_0}^2}$,
$\bar{Z}_6^{\prime} =-{r_6}/{{r_0}^2}$.
Here we have also used $\bar{Z}_{6-}^{\prime}(R)=
\bar{Z}_{2-}^{\prime}(R)=0$.
Using the relations between the inside and
the outside quantities near the shell,
that is Eqs.~(\ref{relation-}) and
(\ref{deltaZ_2})-(\ref{Junction C3}),
we finally obtain the equation of motion of the shell:
\begin{eqnarray}
\label{Final mot of d2d6}
&&A\delta \ddot{R}=B+C\delta R,
\end{eqnarray}
where the definitions of $A, B, C$ are as follows
\begin{eqnarray}
\label{Def of ABC}
A&=&\mu_6V\bar{Z}_2-\mu_2\bar{Z}_6
\\ \nonumber \\
B&=&-\frac{1}{8}\biggl(\frac{\mu_2}{\bar{Z}_2}
+\frac{3\mu_6V}{\bar{Z}_6}\biggl)\delta \Phi_-^{\prime}
-\frac{1}{4}\biggl(\frac{3\mu_2}{\bar{Z}_2}
+\frac{\mu_6V}{\bar{Z}_6}\biggl)\delta \Psi_-^{\prime}
\nonumber \\
&&
+\frac{1}{8r_0^2}\biggl(\frac{\mu_2r_2}{\bar{Z}_2^2}
+\frac{3\mu_6r_6V}{\bar{Z}_6^2}\biggl)
\delta \Phi_{-}
+\frac{3}{64{r_0}^2}
\biggl(\frac{7\mu_2r_2}{\bar{Z}_2^2}
\nonumber \\
&&\;\;\;\;\;\;\;\;
+\frac{3\mu_6r_6V}{\bar{Z}_6^2}
+\frac{3\mu_6r_2V+\mu_2r_6}{\bar{Z}_2\bar{Z}_6}
\biggl)\delta \Psi_{-}
\\ \nonumber \\
C&=&-\frac{1}{128r_0^3}\biggl(\frac{3\mu_2}{\bar{Z}_2}
-\frac{7\mu_6V}{\bar{Z}_6}\biggl)
\biggl(\frac{3r_2}{\bar{Z}_2}+\frac{7r_6}{\bar{Z}_6}
\biggl).
\end{eqnarray}
Here, we have used
$\delta C_{3-}^{\prime}=\delta C_{7-}^{\prime}=0$ inside
the shell. Moreover, using $a:=V/V_{\ast}$ and
$x:={r_{0}}/{|r_{2}|}$, $A$, $B$ and $C$ have the
following simple expressions:
\begin{eqnarray}
\label{Simple Expression of A and C}
&&A=\frac{\mu_{2}}{x} \bigl[(a-1)x-2a\bigr] \\
&&B=\mu_2\Biggl[-\frac{x}{8}\biggl(\frac{1}{x-1}
+\frac{3a}{x+a}\biggl)\delta\Phi_{-}^{\prime}
\nonumber \\
&&\;\;\;\;\;\;\;\;\;\;\;\;\;\;
-\frac{x}{4}\biggl(\frac{3}{x-1}
+\frac{a}{x+a}\biggl)\delta\Psi_{-}^{\prime}
\nonumber \\
&&\;\;\;\;\;\;\;\;\;\;\;\;\;\;
+\frac{1}{8r_{2}}\biggl\{\frac{1}{(x-1)^{2}}
-\frac{3a^{2}}{(x+a)^{2}}\biggl\}\delta\Phi_{-}
\nonumber \\
&&\;\;\;\;\;\;\;\;\;\;\;\;\;\;
+\frac{3}{64r_{2}}\biggl\{\frac{7}{(x-1)^{2}}
-\frac{3a^{2}}{(x+a)^{2}}
\nonumber \\
&&\;\;\;\;\;\;\;\;\;\;\;\;\;\;\;\;\;\;\;\;\;\;
\;\;\;\;\;\;
+\frac{2a}{(x-1)(x+a)}
\biggl\}\delta\Psi_{-}\Biggl] \\
&&C=\frac{\mu_{2}}{128r_{2}^{2}x}
\biggl(\frac{7a}{x+a}
-\frac{3}{x-1}\biggl)^{2}.
\end{eqnarray}
If we take
\begin{eqnarray}
\label{Fourier expansion of R and B}
\delta R&=&\delta \tilde{R}\, e^{i\omega t} \\
B&=&\tilde{B} e^{i\omega t},
\end{eqnarray}
the equation of motion of the shell becomes
\begin{eqnarray}
\label{Fourier Shell}
&&\alpha \delta \tilde{R}=-\tilde{B} \\
&&\alpha=\omega^{2}A+C.
\end{eqnarray}
Therefore, we obtain the amplitude of the forced oscillation:
\begin{eqnarray}
\label{Fourier Shell}
\delta \tilde{R}=-\frac{\tilde{B}}{\alpha}.
\end{eqnarray}

Using the above equation, we can investigate whether
resonances are produced by the forced oscillation or not.
If $\alpha$ can be taken 0, then the amplitude
$\delta \tilde{R}$ becomes infinity and a resonance
is caused. This indicates a kind of ``instability'' of the shell
within our perturbation theory. Furthermore,
{}from the numerical analysis of the Subsec.~\ref{numerical},
the square of the frequency $\omega$ cannot take
negative values. This fact and the behavior of the $\alpha$,
taking notice that $C\geq0$ at any radius, lead us to
classify the region of the radius of the shell into
the following cases.

\noindent (i) When $r_{e}<r_{0}$, $A>0$ (positive
tension) and $C>0$, so that the resonance is not caused.

\noindent (ii) When $r_{0}=r_{e}$, $A=0$ (tensionless)
and $\alpha=C>0$, so that the resonance is not caused.

\noindent (iii) When $r_{0}<r_{e}$, $A<0$ (negative
tension) and $C\geq0$, so that there is necessarily
a real frequency $\omega$ satisfying $\alpha=0$ at
an arbitrary radius of the shell. Therefore, a resonance is
produced and the shell oscillates with huge amplitude at this
radius. In that sense, the shell is ``unstable'' inside the
enhan\c{c}on radius.

{}From the above analysis, the ``unstable'' region of
the shell, where resonances are produced, is $r_0<r_{e}$.
If the shell of the wrapped D6-branes enters the inside
of the enhan\c{c}on radius $r_{e}$, the shell becomes
``unstable" because of the occurrence of the resonances under
small fluctuations of the fields, and the shell may push back
to the enhan\c{c}on radius, where it is stable. This can
explain why the shell is constructed at the enhan\c{c}on
radius.

\subsection{The perturbation of the geometry outside
the shell}
\label{sec.V-iv}

The calculation of the perturbation of the outer
region is tedious but straightforward.
First we examine the matter field equations.
The perturbed equations of the RR fields become
\begin{eqnarray}
\delta{C_{3+}}' & = & -\frac12 (4\delta Z_{2+}
-\delta\Psi_++\delta \Phi_+)\;
{\Cb_3}^{\prime},
\label {dRR3} \\
\delta{C_{7+}}' & = & -\frac12 (4\delta Z_{6+}
+3\delta\Psi_+-3\delta \Phi_+)\;
{\Cb_7}^{\prime}.
\label {dRR7}
\end{eqnarray}
The R-R fields are expressed by the
perturbations of other fields and obtained just by
integrating these equations after we determine other
field.
The perturbed equation of the dilaton field becomes
\begin{eqnarray}
&&-\ztb\zsb\delta \ddot{\Phi}_+
+ \delta \Phi_+^{\prime \prime}
+\frac{2}{r}\delta\Phi_+^{\prime}
\nonumber  \\
&& \;\;\;
=\frac{\phib'}{2}(4\delta \Psi_+
-\delta {Z_{2+}}+3\delta {Z_{6+}})'
\nonumber  \\
& &\;\;\;\;\;\;
+\frac{\phib'}{4}\left(\frac{\ztb'}{\ztb}+\frac{\zsb'}{\zsb}\right)
(\delta \Psi_+-\delta {Z_{2+}}-\delta {Z_{6+}})
\nonumber  \\
& &\;\;\;\;\;\;
-\frac{\ztb^{\prime 2}}{8\ztb^2}
(4\delta \Psi_+ +3\delta {Z_{2+}}-5\delta {Z_{6+}})
\nonumber  \\
& &\;\;\;\;\;\;
+\frac{3\zsb^{\prime 2}}{8\zsb^2}
(4\delta \Psi_+-\delta {Z_{2+}}-\delta {Z_{6+}}).
\label{per-dil}
\end{eqnarray}

Next we turn to the gravitational equations.
The first order perturbation of the Ricci curvature becomes
\begin{eqnarray}
&&\delta R_{00+} = \frac{1}{32\ztb\zsb}
\biggl[8\ztb\zsb(9\delta\ddot{\Psi}_+
-5\delta\ddot{Z}_{2+}
+3\delta\ddot{Z}_{6+})
\nonumber \\
&&\;\;\;\;\;\;
-8(\delta\Psi''_++\delta {Z''_{2+}}
+3\delta {Z''_{6+}}) 
\nonumber \\
&&\;\;\;\;\;\;
- \left(\frac{5\ztb'}{\ztb}+\frac{\zsb'}{\zsb}
+\frac{16}{r}
\right)\delta {Z'_{2+}}
\nonumber \\
&&\;\;\;\;\;\;
+ \left(\frac{15\ztb'}{\ztb}+\frac{3\zsb'}{\zsb}
-\frac{16}{r}
\right)\delta {Z'_{6+}}
\nonumber \\
&&\;\;\;\;\;\;
+ 4\left(\frac{5\ztb'}{\ztb}+\frac{\zsb'}{\zsb}
-\frac{4}{r}
\right)\delta\Psi'_+
\nonumber \\
&&\;\;\;\;\;\;
+ 2\left(\frac{5\ztb''}{\ztb}-\frac{5\ztb^{\prime 2}}{\ztb^2}
+\frac{10}{r}\frac{\ztb'}{\ztb}
+\frac{\zsb''}{\zsb}-\frac{\zsb^{\prime 2}}{\zsb^2} \right.
\nonumber  \\
& &\;\;\;\;\;\;\;\;\;\;\;
+\left.\left. \frac{2}{r}\frac{\zsb'}{\zsb}
\right)(\delta {Z_{2+}}+\delta {Z_{6+}}) \right],
\label{dr00}\\
&&\delta R_{0r+} = \frac{1}{8}
\biggl[ 8(2\delta\dot\Psi'_+
-\delta\dot{Z'}_{2+}+\delta\dot{Z'}_{6+})
\nonumber \\
&&\;\;\;\;\;\;
+\left( \frac{5\ztb'}{\ztb}+\frac{\zsb'}{\zsb}\right)
(\delta\dot{\Psi}_+-\delta\dot{Z}_{2+})
\nonumber  \\
& &\;\;\;\;\;\;\;\;\;\;\;\;
\left.+\left( \frac{3\ztb'}{\ztb}-\frac{\zsb'}{\zsb}\right)
\delta\dot{Z}_{6+}
\right],
\\
&&\delta R_{rr+}=  \frac{1}{32}
\biggl[-\ztb\zsb(\delta\ddot{\Psi}_+
-\delta\ddot{Z}_{2+}-\delta\ddot{Z}_{6+})
\nonumber \\
&&\;\;\;\;\;\;
+8\;(9\delta\Psi''_+-3\delta {Z''_{2+}}
+ 5\delta {Z''_{6+}})
\nonumber \\
&&\;\;\;\;\;\;
-\left(\frac{27\ztb'}{\ztb}-\frac{\zsb'}{\zsb}
-\frac{16}{r}
\right)\delta {Z'_{2+}}
\nonumber \\
&&\;\;\;\;\;\;
-\left(\frac{15\ztb'}{\ztb}
+\frac{35\zsb'}{\zsb}+\frac{16}{r}\zsb'
\right)\delta {Z'_{6+}}
\nonumber \\
&&\;\;\;\;\;\;
\left.-4\left(\frac{3\ztb'}{\ztb}+\frac{9\zsb'}{\zsb}
-\frac{4}{r}
\right)\delta {\Psi'_+}
 \right],
\\
&&\delta R_{mn+}= -\frac{\delta_{mn}}{32\ztb\zsb}
\biggl[8\ztb\zsb(\delta\ddot{\Psi}_+
+\delta\ddot{Z}_{2+}+\delta\ddot{Z}_{6+})
\nonumber \\
&&\;\;\;\;\;\;
-8(\delta\Psi''_++\delta {Z''_{2+}}
+3\delta {Z''_{6+}}) 
\nonumber \\
&&\;\;\;\;\;\;
- \left(\frac{5\ztb'}{\ztb}+\frac{\zsb'}{\zsb}
+\frac{16}{r}
\right)\delta {Z'_{2+}}
\nonumber \\
&&\;\;\;\;\;\;
+ \left(\frac{15\ztb'}{\ztb}+\frac{3\zsb'}{\zsb}
-\frac{16}{r}
\right)\delta {Z'_{6+}}
\nonumber \\
&&\;\;\;\;\;\;
+ 4\left(\frac{5\ztb'}{\ztb}+\frac{\zsb'}{\zsb}
-\frac{4}{r}
\right)\delta\Psi'_+
\nonumber \\
&&\;\;\;\;\;\;
+ 2\left(\frac{5\ztb''}{\ztb}
-\frac{5\ztb^{\prime 2}}{\ztb^2}
+\frac{10}{r}\frac{\ztb'}{\ztb}
+\frac{\zsb''}{\zsb}-\frac{\zsb^{\prime 2}}{\zsb^2}
\right.
\nonumber  \\
& &\;\;\;\;\;\;\;\;\;\;\;\;
+\left.\left. \frac{2}{r}\frac{\zsb'}{\zsb}
\right)(\delta {Z_{2+}}+\delta {Z_{6+}}) \right],
\label{drmm}\\
&&\delta R_{pq+}= \frac{1}{32\zsb}
\biggl[-8\ztb\zsb(\delta\ddot{\Psi}_+
-\delta\ddot{Z}_{2+}+\delta\ddot{Z}_{6+})
\nonumber \\
&&\;\;\;\;\;\;
+8(\delta\Psi''_+-\delta {Z''_{2+}}
+\delta {Z''_{6+}}) 
\nonumber \\
&&\;\;\;\;\;\;
+ \left(\frac{12\ztb'}{\ztb}+\frac{\zsb'}{\zsb}
-\frac{16}{r}
\right)\delta {Z'_{2+}}
\nonumber \\
&&\;\;\;\;\;\;
- \left(\frac{3\ztb'}{\ztb}+\frac{3\zsb'}{\zsb}
+\frac{16}{r}
\right)\delta {Z'_{6+}}
\nonumber \\
&&\;\;\;\;\;\;
+ \left(\frac{9\ztb'}{\ztb}-\frac{4\zsb'}{\zsb}
+\frac{16}{r}
\right)\delta\Psi'_+
\nonumber \\
&&\;\;\;\;\;\;
+ 2\left(\frac{3\ztb''}{\ztb}
-\frac{3\ztb^{\prime 2}}{\ztb^2}
+\frac{6}{r}\frac{\ztb'}{\ztb}
-\frac{\zsb''}{\zsb}
-\frac{2\zsb^{\prime 2}}{\zsb^2} \right.
\nonumber  \\
& &\;\;\;\;\;\;\;\;\;\;\;\;
-\left.\left. \frac{2}{r}\frac{\zsb'}{\zsb}
\right)\delta {Z_{6+}} \right]
\delta_{pq},
\label{drpq}
\end{eqnarray}

On the other hand, the energy momentum tensor
is perturbed as
\begin{eqnarray}
&&\delta T_{00+} = \frac{1}{64\ztb\zsb}
\biggl[
\nonumber \\
&& \;\;\;\;\;\;\;
-\left(\frac{57\ztb^{\prime 2}}{\ztb^2}
-\frac{6\ztb^{\prime}\zsb^{\prime}}{\ztb\zsb}
+\frac{33\zsb^{\prime 2}}{\zsb^2}\right)\delta Z_{2+}
\nonumber \\
&& \;\;\;\;\;\;\;
+
\left(\frac{7\ztb^{\prime 2}}{\ztb^2}
+\frac{6\ztb^{\prime}\zsb^{\prime}}{\ztb\zsb}
-\frac{33\zsb^{\prime 2}}{\zsb^2}\right)\delta Z_{6+}
\nonumber \\
&&\;\;\;\;\;\;\;
+8\left(\frac{5\ztb^{\prime 2}}{\ztb^2}
+\frac{\zsb^{\prime 2}}{\zsb^2}\right)\delta \Psi_+
-8\left(\frac{\ztb^{\prime 2}}{\ztb^2}
-\frac{3\zsb^{\prime 2}}{\zsb^2}\right)\delta \Phi_+
\nonumber \\
&&\;\;\;\;\;\;\;
\left.
+8\left(\frac{\ztb^{\prime}}{\ztb}
-\frac{3\zsb^{\prime}}{\zsb}\right)\delta \Phi'_+\right]
\\
\nonumber \\
&&\delta T_{0r+} = \frac18
\left(\frac{\ztb^{\prime}}{\ztb}
-\frac{3\zsb^{\prime}}{\zsb}\right)
\delta \dot{\Phi}_+, \\
\nonumber \\
&&\delta T_{rr+} = \frac{1}{8}
\left[\left(\frac{5\ztb^{\prime 2}}{\ztb^2}
+\frac{\zsb^{\prime 2}}{\zsb^2}\right)\delta Z_{2+}
\right.
\nonumber \\
&&\;\;\;\;\;\;\;
-\left(\frac{3\ztb^{\prime 2}}{\ztb^2}
-\frac{\zsb^{\prime 2}}{\zsb^2}\right)\delta Z_{6+}
-\left(\frac{5\ztb^{\prime 2}}{\ztb^2}
+\frac{\zsb^{\prime 2}}{\zsb^2}\right)\delta \Psi_{+}
\nonumber \\
&&\;\;\;\;\;\;\;
+
\left(\frac{\ztb^{\prime 2}}{\ztb^2}
-\frac{3\zsb^{\prime 2}}{\zsb^2}\right)\delta \Phi_{+}
\left.
+\left(\frac{\ztb^{\prime}}{\ztb}
-\frac{3\zsb^{\prime}}{\zsb}\right)\delta \Phi'_+\right],
\nonumber \\
\\
\nonumber \\
&&\delta T_{mn+} = -\delta T_{00+}\delta_{mn}, \\
\nonumber \\
&&\delta T_{pq+} = \frac{1}{64\zsb}
\left[-8\left(\frac{5\ztb^{\prime 2}}{\ztb^2}
+\frac{\zsb^{\prime 2}}{\zsb^2}\right)\delta Z_{2+}
\right.
\nonumber \\
&&\;\;\;\;\;\;\;
+3\left(\frac{3\ztb^{\prime 2}}{\ztb^2}
-\frac{2\ztb^{\prime}\zsb^{\prime}}{\ztb\zsb}
-\frac{5\zsb^{\prime 2}}{\zsb^2}\right)\delta Z_{6+}
\nonumber \\
&&\;\;\;\;\;\;\;
+8\left(\frac{5\ztb^{\prime 2}}{\ztb^2}
+\frac{\zsb^{\prime 2}}{\zsb^2}\right)\delta \Psi_+
-8\left(\frac{\ztb^{\prime 2}}{\ztb^2}
-\frac{3\zsb^{\prime 2}}{\zsb^2}\right)\delta \Phi_+
\nonumber \\
&&\;\;\;\;\;\;\;
\left.
-8\left(\frac{\ztb^{\prime 2}}{\ztb^2}
-\frac{3\zsb^{\prime 2}}{\zsb^2}\right)\delta \Phi'_+
\right]
\delta_{pq}.
\end{eqnarray}
Here we have used Eqs.~(\ref{dRR3}) and (\ref{dRR7}).

Since
$\delta T_{00+}+\delta T_{mm+}=0$,
combining the $(00)$ and
$(mm)$ components of the Einstein equations gives
the following useful relation
\begin{equation}
\delta\ddot{\Psi}_+ = \frac34 \delta\ddot{Z}_{2+}
- \frac14 \delta\ddot{Z}_{6+}.
\end{equation}
By assuming the time dependence of these perturbed
variables
as harmonic oscillator as
Eqs.~(\ref{decouple1})-(\ref{decouple4}),
we find
\begin{equation}
\xi_+ = \frac34 \zeta_{2+} - \frac14 \zeta_{6+}.
\label{rel}
\end{equation}

By the $(0, 0)$, $(r, r)$ and $(p, q)$ components of
the perturbed Einstein equations,
we find the equations of $\zeta_{2+}$ and $\zeta_{6+}$,
\begin{eqnarray}
\label{z2-out}
{\zeta_{2+}}^{\prime \prime}
&-&\left(\frac{\ztb'}{\ztb}-\frac{2}{r} \right)
\zeta_{2+}^{\prime}
-\frac{\ztb'}{\ztb} \zeta_{+6}^{\prime}
%
-\frac18\left(\frac{5\ztb''}{\ztb}+\frac{\zsb''}{\zsb}
\right.
\nonumber \\
&+&\frac{5\ztb^{\prime 2}}{\ztb^2}
+\frac{\zsb^{\prime 2}}{\zsb^2}
+\frac{10}{r}\frac{\ztb'}{\ztb}
%
\left. +\frac{2}{r}\frac{\zsb'}{\zsb}
\right) \zeta_{2+}
\nonumber \\
&-&\frac18\left(\frac{8\ztb''}{\ztb}
-\frac{7\ztb^{\prime 2}}{\ztb^2}
+\frac{13\zsb^{\prime 2}}{\zsb^2}
+\frac{16}{r}\frac{\ztb'}{\ztb}
\right) \zeta_{6+}
\nonumber \\
&-&\frac{1}{2}\left(\frac{\ztb^{\prime 2}}{\ztb^2}
-3\frac{\zsb^{\prime 2}}{\zsb^2}\right) \eta_+
+\omega^2 \ztb \zsb \zeta_{2+}=0,
\nonumber \\
\end{eqnarray}
\begin{eqnarray}
{\zeta_{6+}}^{\prime \prime}
& +&\frac13\left(\frac{2\ztb'}{\ztb}
-\frac{\zsb'}{\zsb}\right)
\zeta_{2+}^{\prime}
+\frac13\left(\frac{2\ztb'}{\ztb} -\frac{\zsb'}{\zsb}
\right.
\nonumber \\
&& \left.
+\frac6r\right)
\zeta_{6+}^{\prime}
-\frac{1}{24}\left(\frac{5\ztb''}{\ztb}
+\frac{\zsb''}{\zsb}
-\frac{10\ztb^{\prime 2}}{\ztb^2}
-\frac{6\zsb^{\prime 2}}{\zsb^2}
\right.
\nonumber \\
& & \left.
+\frac{10}{r}\frac{\ztb'}{\ztb}
+\frac{2}{r}\frac{\zsb'}{\zsb}
\right) \zeta_{2+}
+\frac{1}{3}\left(\frac{2\ztb''}{\ztb}-\frac{\zsb''}{\zsb}
\right.
\nonumber \\
&& \left.
-\frac{2\ztb^{\prime 2}}{\ztb^2}
+\frac{\zsb^{\prime 2}}{\zsb^2}+\frac{4}{r}\frac{\ztb'}{\ztb}
+\frac{2}{r}\frac{\zsb'}{\zsb}
\right) \zeta_{6+}
\nonumber \\
&+&\frac{1}{3}\left(\frac{\ztb^{\prime 2}}{\ztb^2}
-9\frac{\zsb^{\prime 2}}{\zsb^2}\right) \eta_+
+\omega^2 \ztb \zsb \zeta_{6+} =0.
\nonumber \\
\end{eqnarray}
By using the relation (\ref{rel}), the dilaton field equation
becomes
\begin{eqnarray}
& &\eta_+''+\frac{2}{r}\eta_+'
=
\phib'({\zeta_{2+}}'+{\zeta_{6+}}')
\nonumber  \\
& & \;\;\;\;\;\;\;
-\frac{\phib'}{16}\left(\frac{\ztb'}{\ztb}+\frac{\zsb'}{\zsb}\right)
({\zeta_2}+5\,{\zeta_{6+}})
\nonumber  \\
& & \;\;\;\;\;\;\;
-\frac{3}{4}\left(\frac{\ztb^{\prime 2}}{\ztb^2}
-\frac{\zsb^{\prime 2}}{\zsb^2}\right)
({\zeta_{2+}}-{\zeta_{6+}})
-\omega^2 \ztb \zsb \eta_+.
\nonumber \\
\label{dil-out}
\end{eqnarray}

\subsection{Numerical analysis}
\label{numerical}

Now, everything has been prepared for the perturbation analysis,
i.e., the perturbed solutions of the geometry inside the shell
(\ref{relation-}) and (\ref{EinM-S}),
the matching and the junction conditions
(\ref{junc-1}) and (\ref{deltaZ_2})-(\ref{delta-Phi}),
the equation of motion for the shell (\ref{Final mot of d2d6}),
and the perturbed equations in the outer geometry
(\ref{rel})-(\ref{dil-out}).
Since we know the solutions inside the shell, we can integrate
{}from the $r_0$ to outside region with the boundary
value and the junction conditions of the
perturbed functions at the shell.
If there is an eigenmode with
a negative eigenvalue $\omega^2$, the
background solution is unstable against the perturbations.

Our numerical calculations show, surprisingly,
that no eigenmodes with negative $\omega^2$ can be
found for any radii~($r_0>|r_2|$) of the shell.
Furthermore, as shown in Subsec.~\ref{sec.V-iii}, we found
resonances for the case (iii). The followings are
two typical examples for the numerical calculations.

\begin{figure}
\centerline{\epsfxsize=8.0cm \epsfbox{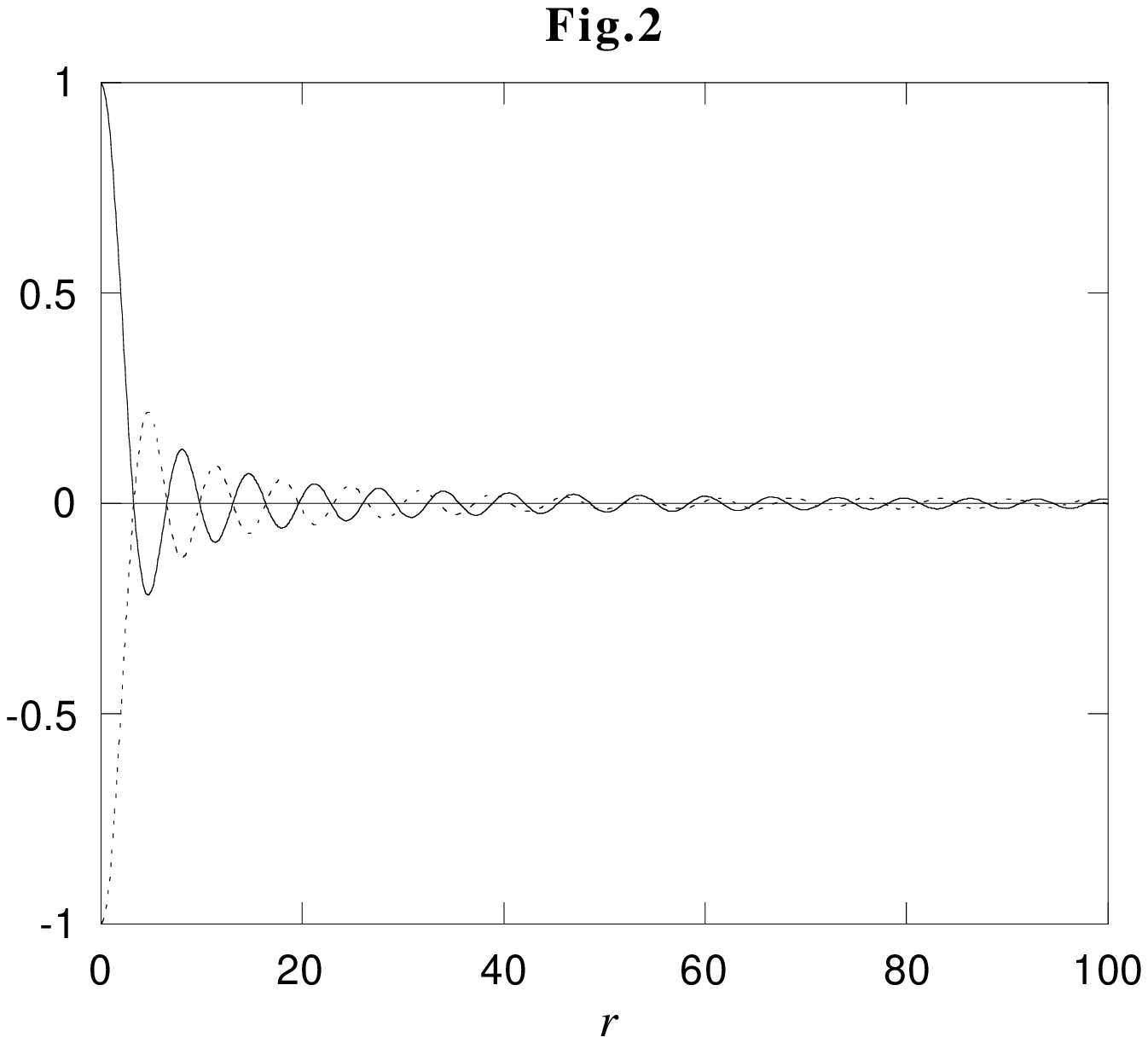}}
\caption{The typical configuration of the perturbed
functions with the parameter $g=0.1$, $N=100$,
$V=3000$. The solid and the dotted lines are
$\delta Z_2$ and $\delta Z_6$, respectively.
The enhan\c{c}on radius of these parameters
is $r_e=10.8123$.
The radius of the shell is $r_0=20.0$
and the eigenvalue is $\omega^2=1.0$.
}
\protect
\label{stable}
\end{figure}

Fig.~\ref{stable} shows configurations of the stable modes
when the shell is located at $r=r_0>r_e$. The
enhan\c{c}on radius of the parameters in Fig.~\ref{stable}
is $r_e=10.8123$. Outside of the
shell, the perturbed functions couple each other
complicatedly and they show disordered behaviors.
The stable modes show a continuous spectrum as usual.

\begin{figure}
\centerline{\epsfxsize=8.0cm \epsfbox{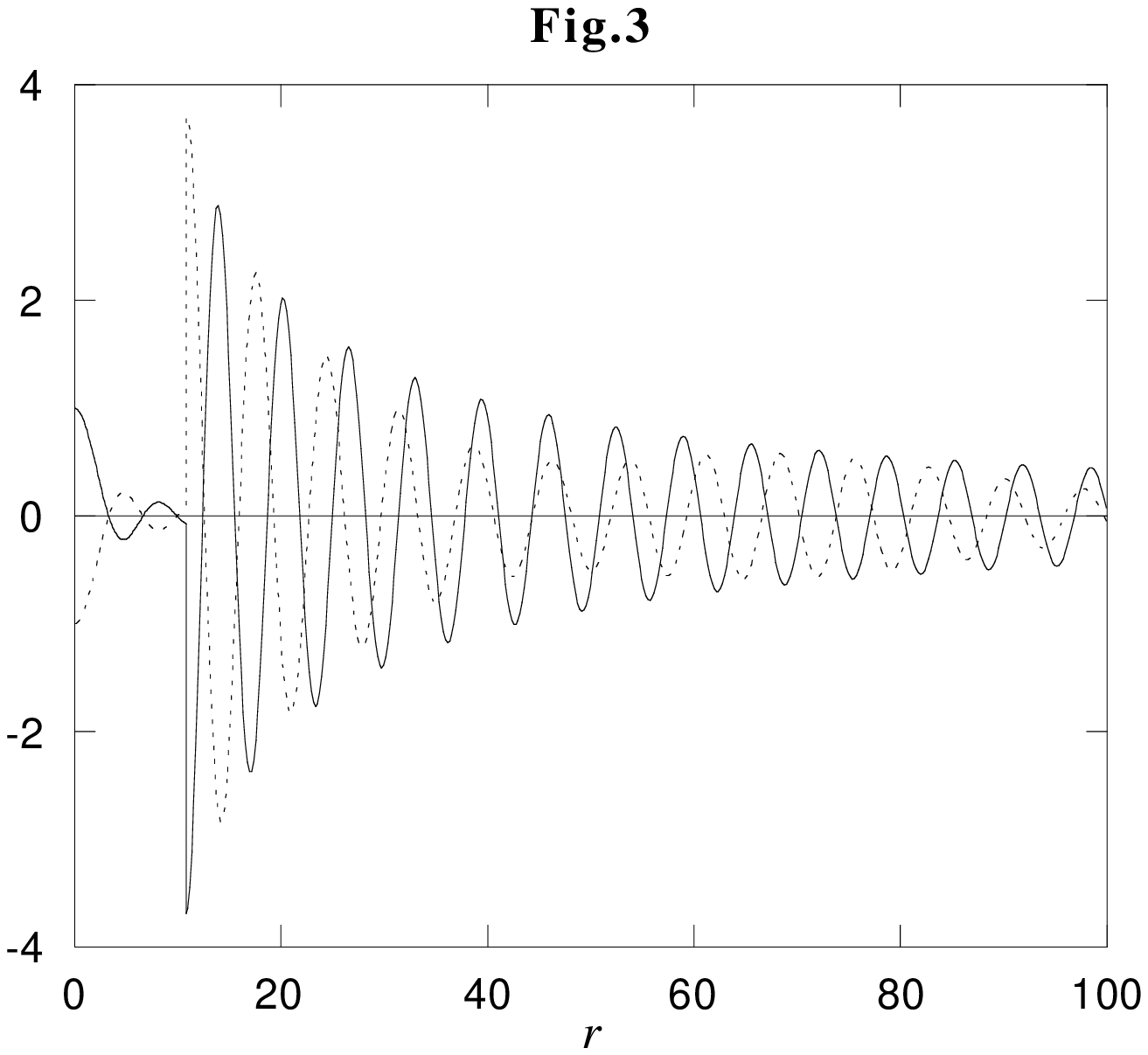}}
\caption{The typical configuration of the perturbed
functions with the
parameter $g=0.1$,
$N=100$,
$V=3000$.
The radius of the shell is $r_0=10.8$
and the eigenvalue is $\omega^2=1.0$.
The solid and the dotted lines are
$\delta Z_2$ and $\delta Z_6$, respectively.
Since the shell locates near the resonance radius
$r_{reso}=10.8071$
the amplitudes of the perturbed functions
become very large at the shell.}
\protect
\label{reso}
\end{figure}

When the shell is located inside of the enhan\c{c}on radius,
the configurations of the perturbed functions are similar
to the previous case $r_0>r_e$ except for the existence of
the resonance radii. For the same parameter in
Fig.~\ref{stable}, the resonance radii are
$r_{reso}=2.74099$, $10.8071$ in $|r_{2}|<r_{reso}<r_{e}$,
where $|r_{2}|=2.59758$ and $r_e=10.8123$.
Fig.~\ref{reso} shows configurations when the shell is
located near the resonance radius. The perturbed 
functions becomes very large at the shell by the
resonance.

When the shell crosses the enhan\c{c}on radius,
the total tension of the shell becomes negative.
So we intuitively guess that the shell has eigenmodes
with negative $\omega^2$.
Our numerical calculations show, however,
that there is no such eigenmode.
Since the shell we are considering couples to the bulk
fields, it is considered that the lost of the energy by
moving the shell is compensated by the increase of
the bulk fields energy.

\vspace{1cm}

\section{Conclusion and discussions}
\label{sec.VI}
We have investigated the stability of the shell of wrapped D6-branes
on K3 from the point of view of supergravity.
As shown in Sec.~\ref{sec.II}, if a probe D6-brane (wrapped on K3)
is kicked inward
slightly at the outside of the enhan\c{c}on radius $r=r_e$, the velocity
reaches the speed of light at the radius.
This behavior seems to be in contradiction to the behavior of
the enhan\c{c}on mechanism, which states
that the D6-branes shell sits around the enhan\c{c}on radius.
So, what mechanism stabilizes the non-probe D6-brane?

In the probe case, the background spacetime is fixed as a
supersymmetric D6-D2 brane supergravity solution. Therefore,
there is no obstacle to stop the motion of the probe since
no force acts on it.
In the non-probe case, the background spacetime gets excited by the
motion of the shell and becomes non-supersymmetric.
Then, new forces act on the D6 branes shell, as seen in Eq.~(\ref{Final
mot of d2d6}).
In other words, the kinematic energy of the
shell could be lost through the waves such as gravitational and the
dilaton waves.
This is one of the reasons why the non-probe D6-branes shell is stable.

Quite interestingly, it is found numerically that there is
no eigenmode with negative $\omega^2$ for any radius of the shell.
One may doubt it when the shell has negative tension, because its
energy falls as it moves. This naive picture is not applicable
to our case because the bulk energy is also changed.
For example, let us consider the shell located inside of
the enhan\c{c}on radius. Because of the negative tension,
the energy $E$ falls as $E\propto - (\delta\dot{R})^2$
when it moves slowly. On the other hand, the bulk energy
of the dilaton field and the R-R fields may increase as the shell
moves. So, we cannot discuss the stability of the whole system
in terms of $E$ only. Indeed, the D6-D2 system under consideration
provides a notable example for demonstrating that D-branes shell
with negative tension is stable in the sense that there is
no eigenmode with negative $\omega^2$.

In Sec. V, we found that there is a resonant frequency mode whenever
the shell is located at $r_0<r_e$. Because every time-dependent
function $f(t)$ contains this mode as a Fourier mode, the shell would
oscillate with huge amplitude. In this sense the shell is
``unstable'' generically whenever the shell is located
inside the enhan\c{c}on radius.
This fact suggests that the shell located inside of
the enhan\c{c}on radius is pushed back to the enhan\c{c}on radius,
where it is stable.
Therefore, we may say that {\it the enhan\c{c}on radius is
naturally selected as a suitable position of the wrapped D6-branes
shell even from the viewpoint of the ``classical''
supergravity with the D-branes.}
This coincides with the enhan\c{c}on picture stated in the
introduction.

 It is interesting to investigate the stability of other models such
as brane world scenario~\cite{RS}. There is a common belief that
the brane with negative tension is unstable against the perturbation.
By analogy with the present analysis, we may expect that it can be
stabilized if one considers the perturbation of the whole system
including the self gravity of the brane.
It is also interesting to investigate the non-commutative effect of
N D6-branes or non-linear effect of the D6-D2 branes system.
After such steady works, we could see new physics of gravity
combined with string theory.

\section*{Acknowledgement}
We would like to thank Gary T. Horowitz for useful discussions.
We also thank Makoto Natsuume and Justin R. David for reading
the manuscript and making a number of helpful suggestions.
K. M. is supported by a JSPS Postdoctoral Fellowship for Research
Abroad and S. Y. is supported by Rikkyo University
Grant for the Promotion of Research.


\begin{references}
\bibitem{B}
K. Behrndt, Nucl. Phys. {\bf B455}, 188 (1995).
\bibitem{KL}
R. Kallosh and A. Linde,
Phys. Rev. D{\bf 52}, 7137 (1995).
\bibitem{CY}
M. Cveti\v{c} and D. Youm,
Phys. Lett. B {\bf 359}, 87 (1995).
\bibitem{P}
R. Ponrose, Riv. Nuovo Cimento {\bf 1}, 252 (1969)
\bibitem{JPP}
C. V. Johnson, A. W. Peet, and J. Polchinski,
Phys. Rev. D{\bf 61}, 086001 (2000).
\bibitem{J}
C. V. Johnson,
Phys. Rev. D{\bf 63}, 065004 (2001).
\bibitem{BVS}
M. Bershadsky, C. Vafa, and V. Sadov, Nucl. Phys. {\bf B463}, 398 (1996).
\bibitem{GHM}
M. Green, J. A. Harvey, and G. Moore, Class. Quantum Grav. {\bf 14}, 47 (1997).
\bibitem{DJM}K. Dasgupta, D. P. Jatkar, and S. Mukhi,
Nucl. Phys. {\bf 523}, 465 (1998)
\bibitem{BBG}
C.~P.~Bachas, P. Bain and M.~B.~Green, JHEP {\bf05},
011 (1999)
\bibitem{Israel}
W. Israel, Nuovo Cimento {\bf 44B}, 1 (1966)
\bibitem{Talk}
K.~Maeda, Talk presented at {\it 17th Pacific Coarst Gravity\\
 Meeting},\hspace{2mm}\noindent Institute of Theoretical Physics, UCSB,
Santa Barbara, March, 2001, \\
http://www.itp.ucsb.edu/online/pcgm17.
\bibitem{JMPR}
C. V. Johnson, R. C. Myers, A. W. Peet, and S. F. Ross,
hep-th/0105077.
\bibitem{HE}
S.~W.~Hawking and G.~F.~R.~Ellis, {\it The large scale structure
of space-time}, Cambridge U. Press, Cambridge, 1973.
\bibitem{com1}
T. Torii, K. Maeda, M. Narita and S. Yahikozawa,
{\it Proceeding of the Meeting on Frontier of  Cosmology and Gravitation},
to be published.
\bibitem{RS}
L. Randall and R. Sundrum, Phys. Rev. Lett. {\bf 83}, 3370 (1999);
{\bf 83} 4690 (1999).

\end{references}
\end{document}